\documentclass[acmsmall,nonacm]{acmart}

\renewcommand\footnotetextcopyrightpermission[1]{} 

\AtBeginDocument{%
	\providecommand\BibTeX{{%
			\normalfont B\kern-0.5em{\scshape i\kern-0.25em b}\kern-0.8em\TeX}}}

\newcommand{\quo}[1]{``{#1}''}
\newcommand{\highlightI}[1]{\textit{#1}}
\newcommand{\highlightB}[1]{\textbf{#1}}

\usepackage{hyperref}
\usepackage{array}
\usepackage{booktabs}

\usepackage{color, colortbl}
\definecolor{lightgray}{gray}{0.9}

\usepackage{pgffor}

\usepackage{amsmath}
\newcommand{\epv}{E}
\newcommand{\real}{\mathbb{R}}

\newcommand{\mathcomma}{,\;\;\;}
\newcommand{\norm}[1]{\left\lVert#1\right\rVert}
\newcommand{\argmin}{\mathop{\mathrm{argmin}}}

\usepackage{pdfrender}
\newcommand*{\boldcheckmark}{
	\textpdfrender{
		TextRenderingMode=FillStroke,
		LineWidth=1pt, 
	}{\checkmark}
}
\newcommand*{\dashcheckmark}{
	\textpdfrender{
		TextRenderingMode=Stroke,
		LineWidth=0.5pt,
		LineDashPattern=[2 2]0,
	}{\checkmark}
}

\begin{document}
	
	\title[Smell Pittsburgh]{Smell Pittsburgh: Engaging Community Citizen Science for Air Quality}
	
	\author{Yen-Chia Hsu}
	\affiliation{\institution{Carnegie Mellon University}\country{USA}}
	\email{yenchiah@andrew.cmu.edu}
	
	\author{Jennifer Cross}
	\affiliation{\institution{Tufts University}\country{USA}}
	\email{jennifer.cross@tufts.edu}
	
	\author{Paul Dille}
	\affiliation{\institution{Carnegie Mellon University}\country{USA}}
	\email{pdille@andrew.cmu.edu}
	
	\author{Michael Tasota}
	\affiliation{\institution{Carnegie Mellon University}\country{USA}}
	\email{tasota@andrew.cmu.edu}
	
	\author{Beatrice Dias}
	\affiliation{\institution{Carnegie Mellon University}\country{USA}}
	\email{mdias@andrew.cmu.edu}
	
	\author{Randy Sargent}
	\affiliation{\institution{Carnegie Mellon University}\country{USA}}
	\email{rsargent@andrew.cmu.edu}
	
	\author{Ting-Hao (Kenneth) Huang}
	\affiliation{\institution{Pennsylvania State University}\country{USA}}
	\email{txh710@psu.edu}
	
	\author{Illah Nourbakhsh}
	\affiliation{\institution{Carnegie Mellon University}\country{USA}}
	\email{illah@andrew.cmu.edu}
	
	\renewcommand{\shortauthors}{Hsu et al.}
	
	\begin{abstract}
		Urban air pollution has been linked to various human health concerns, including cardiopulmonary diseases. Communities who suffer from poor air quality often rely on experts to identify pollution sources due to the lack of accessible tools. Taking this into account, we developed \highlightB{\highlightI{Smell Pittsburgh}}, a system that enables community members to report odors and track where these odors are frequently concentrated. All smell report data are publicly accessible online. These reports are also sent to the local health department and visualized on a map along with air quality data from monitoring stations. This visualization provides a comprehensive overview of the local pollution landscape. Additionally, with these reports and air quality data, we developed a model to predict upcoming smell events and send push notifications to inform communities. We also applied regression analysis to identify statistically significant effects of push notifications on user engagement. Our evaluation of this system demonstrates that engaging residents in documenting their experiences with pollution odors can help identify local air pollution patterns, and can empower communities to advocate for better air quality. All citizen-contributed smell data are publicly accessible and can be downloaded from \textit{\textbf{\url{https://smellpgh.org}}}.
	\end{abstract}

	
	
	
	\maketitle
	
	\section{Introduction}

Air pollution has been associated with adverse impacts on human health, including respiratory and cardiovascular diseases~\cite{Kampa-2008, Pope-2006, Dockery-1993, WHO-2016, WHO-2016-air-quality}. Addressing air pollution often involves negotiations between corporations and regulators, who hold power to improve air quality. However, the communities, who are directly affected by the pollution, are rarely influential in policy-making. Their voices typically fail to persuade decision-makers because collecting and presenting reliable evidence to support their arguments is resource-intensive. Forming such evidence requires collecting and analyzing multiple sources of data over a large geographic area and an extended period. This task is challenging due to the requirements of financial resources, organizational networks, and access to technology. Due to power imbalance and resource inequality, affected residents usually rely on experts in governmental agencies, academic institutions, or non-governmental organizations to analyze and track pollution sources.

\begin{figure*}[t]
	\centering
	\includegraphics[width=1\columnwidth]{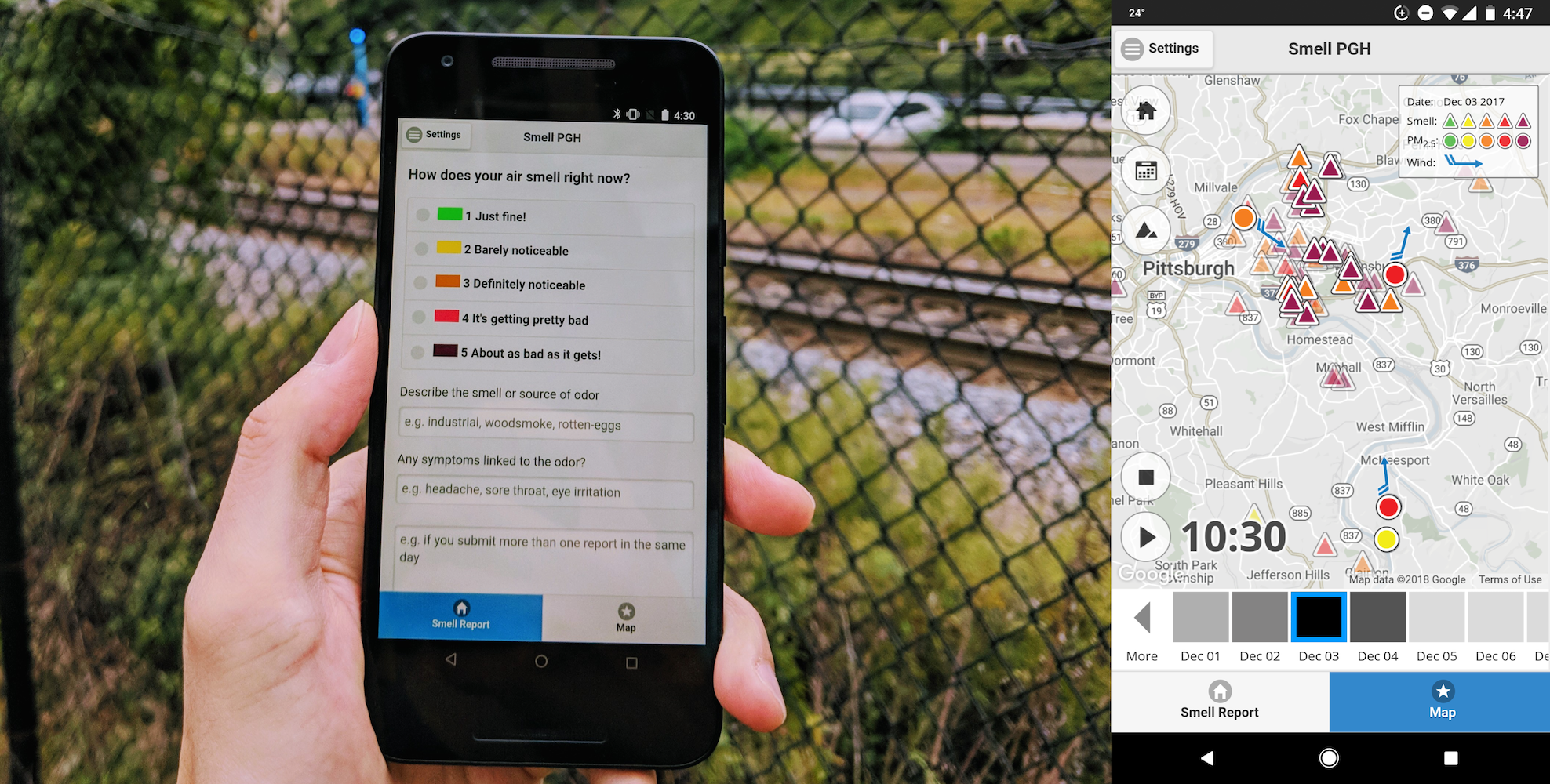}
	\caption{The user interface of Smell Pittsburgh. The left image shows the submission console for describing smell characteristics, explaining symptoms, and providing notes for the local health department. The right image shows the visualization of smell reports, sensors, and wind directions.}
	\label{fig:UI_smell_pittsburgh}
\end{figure*}

A straightforward solution is to empower the affected communities directly. In this research, we demonstrate how citizen science can be used for communities to pool resources and gather evidence for advocacy. Data-driven evidence, especially when integrated with narratives, is essential for communities to make sense of local environmental issues and take action~\cite{Ottinger-2017-sense}. However, citizen-contributed data is often held in low regard because the information can be unreliable or include errors during data entry. Also, sufficient citizen participation and data transparency are required for the evidence to be influential. For instance, the city involved in this study, Pittsburgh, is one of the most polluted cities in the United States~\cite{AmericanLungAssociation}. Currently, Pittsburgh citizens report air quality problems to the local health department via its phone line or website.

Nevertheless, the quality of the gathered data is doubtful. Citizens may not remember the exact time and location that pollution odors occurred. Asking citizens to submit complaints retrospectively is hard for capturing accurate details and prone to errors. Such errors can result in missing or incomplete data that can affect the outcome of statistical analysis to identify pollution sources~\cite{Devillers-2006}. Furthermore, the reporting process is not transparent and does not encourage citizens to contribute data. There is no real-time feedback or ways of sharing experiences to forge a sense of community. Without data that adequately represents the community, it is difficult to know if an air pollution problem is at a neighborhood or city-wide scale. This approach is inadequate for data collection and hinders participation in bringing air quality issues to the attention of regulators and advocating for policy changes.

Because of these challenges, resident-reported smell data did not gain much attention as a critical tool for monitoring air pollution. However, literature has shown that the human olfactory system can distinguish more than one trillion odors~\cite{Bushdid-2014} and outperform sensitive measuring equipment in odor detection tasks~\cite{Shepherd-2004}. Although there have been discussions about the potential of using smell to indicate pollution events and support decision making~\cite{Ottinger-2010, Obrist-2014}, no prior works collected long-term smell data at a city-wide scale and studied if these data are useful for air pollution monitoring and community advocacy.

We propose a system, \highlightB{\highlightI{Smell Pittsburgh}} \cite{smell-pgh-website}, for citizens to report pollution odors to the local health department with accurate time and GPS location data via smartphones. The system visualizes odor complaints in real-time, which enables residents to confirm their experiences by viewing if others also share similar experiences. Additionally, we present a dataset of smell reports and air quality measurements from nearby monitoring stations over 21 months~\cite{smell-dataset-2018}. We use the dataset to develop a model that predicts upcoming pollution odors and send push notifications to users. We also apply machine learning to identify relationships between smell reports and air quality measurements. Moreover, we analyze the effects of different types of push notifications to determine if they are related to increases in user engagement. Finally, we describe qualitative and quantitative studies for understanding changes in user engagement and motivation. To the best of our knowledge, Smell Pittsburgh is the first system of its kind that demonstrates the potential of collecting and using smell data to form evidence about air quality issues at a city-wide scale. Although stakeholders typically view odor experiences as subjective and noisy, our work shows that smell data is beneficial in identifying urban air pollution patterns and empowering communities to pursue a sustainable environment. All citizen-contributed smell data are publicly accessible and can be downloaded from the project website~\cite{smell-pgh-website}.
	
	\section{Related Work}

This research is rooted in citizen science, which empowers amateurs and professionals to form partnerships and produce scientific knowledge \cite{EU-2013, Bonney-2014, Bonney-2016, McKinley-2015, Eitzel-2017}. Historically, there exist both research and community-oriented strategies~\cite{Cooper-2016}. Research-oriented citizen science aims to address large-scale research questions which are infeasible for scientists to tackle alone \cite{Bonney-Ballard-2009, Silvertown-2009, Cohn-2008, Dickinson-Shirk-2012, Dickinson-Bonney-2012, Miller-Rushing-2012, Bonney-Cooper-2009, Cooper-2007}. Research questions under this strategy are often driven by professional scientists. Researchers applying this strategy study how scientists can encourage the public to participate in scientific research. In contrast, \highlightB{community-oriented citizen science} aims to democratize science by equipping citizens with tools to directly target community concerns for advocacy~\cite{Irwin-1995, Greaves-1980, Wilsdon-2005, Stilgoe-2009, Irwin-2001, Paulos-2008, Irwin-2006, Stilgoe-2014, Ottinger-2016, Chari-2017, Hsu-2018-thesis, Corburn-2005}. Research questions under this strategy are often driven by community members, exploring how scientists can engage in social and ethical issues that are raised by citizens or communities. Our research focuses on the community-oriented approach, defined as the \highlightB{Community Citizen Science (CCS)} framework~\cite{Hsu-2019-CCS}. CCS is highly related to sustainable Human-Computer Interaction~\cite{DiSalvo-2009, DiSalvo-Sengers-2010, Brynjarsdottir-2012, Blevis-2007, Mankoff-2007, Dourish-2010}, which studies the intervention of information technology for increasing the awareness of sustainability, changing user behaviors, and influencing attitudes of affected communities. We seek to generate scientific knowledge from community data to support citizen-driven exploration, understanding, and dissemination of local air quality concerns.

\subsection{Community Data in Citizen Science}

Modern technology allows communities to collect data that can contextualize and express their concerns. There are typically two types of community data, which are generated from either sensors or proactive human reports. Each type of data provides a small fragment of evidence. When it comes to resolving and revealing community concerns, human-reported data can show how experiences of residents are affected by local issues, but it is typically noisy, ambiguous, and hard to quantify at a consistent scale. Sensing data can complement human-reported data by providing temporally dense and reliable measurements of environmental phenomena but fails to explain how these phenomena affect communities. Without integrating both types of data, it is difficult to understand the context of local concerns and produce convincing evidence.

\subsubsection{Human-Reported Data}
Human-reported data includes observations contributed by users. Modern computational tools can collect volunteered geographic information~\cite{Haklay-2013} and aggregate them to produce scientific knowledge. However, most prior works focused on collecting general information of particular interest, rather than data of a particular type of human sense, such as odor. \highlightI{Ushahidi} gathers crisis information via text messages or its website to provide timely transparent information to a broader audience~\cite{Okolloh-2009}. \highlightI{Creek Watch} is a monitoring system which enabled citizens to report water flow and trash data in creeks~\cite{Kim-Robson-2011}. \highlightI{Sensr} is a tool for creating environmental data collection and management applications on mobile devices without programming skills~\cite{Kim-Mankoff-2013, Kim-2015}. \highlightI{Encyclopedia of Life} is a platform for curating species information contributed by professionals and non-expert volunteers~\cite{Rotman-Procita-2012}. \highlightI{eBird} is a crowdsourcing platform to engage birdwatchers, scientists, and policy-makers to collect and analyze bird data collaboratively~\cite{Sullivan-2014, Sullivan-2009}. \highlightI{Tiramisu} was a transit information system for collecting GPS location data and problem reports from bus commuters~\cite{Zimmerman-2011}. One of the few examples focusing on information of a specific modality is \highlightI{NoiseTube}, a mobile application that empowered citizens to report \highlightI{noise} via their mobile phones and mapped urban noise pollution on a geographical heatmap~\cite{Maisonneuve-2009, Dhondt-2012}. The tool could be utilized for not only understanding the context of urban noise pollution but also measuring short-term or long-term personal exposure.

\subsubsection{Sensing Data}
Sensing data involves environmental measurements quantified with sensing devices or systems, which enable citizens to monitor their surroundings with minimal to no assistance from experts. However, while many prior works used sensors to monitor air pollution, none of them complemented the sensing data with human-reported data. \highlightI{MyPart} is a low-cost and calibrated wearable sensor for measuring and visualizing airborne particles~\cite{Tian-2016}. \highlightI{Speck} is an indoor air quality sensor for measuring and visualizing fine particulate matter~\cite{Taylor-2015,Taylor-2016}. Kim {\em et al.} implemented an indoor air quality monitoring system to gather air quality data from commercial sensors~\cite{Kim-Paulos-2013}. Kuznetsov {\em et al.} developed multiple air pollution monitoring systems which involved low-cost air quality sensors and a map-based visualization~\cite{Kuznetsov-2011, Kuznetsov-2014}. Insights from these works showed that sensing data, especially accompanied by visualizations, could provide context and evidence that might raise awareness and engage local communities to participate in political activism. But none of these work asked users to report odors, and thus can not directly capture how air pollution affects the living quality of community members.


\subsection{Machine Learning for Citizen Science}

Citizen science data are typically high-dimensional, noisy, potentially correlated, and spatially or temporally sparse. The collected data may also suffer from many types of bias and error that sometimes can even be unavoidable~\cite{Budde-2017, Bird-2014}. Making sense of such noisy data has been a significant concern in citizen science~\cite{Ottinger-2017-crowd, Newman-2012}, especially for untrained contributors~\cite{Cohn-2008, Ottinger-2010, Bonney-2014, Broeder-2016}. To assist community members in identifying evidence from large datasets efficiently, prior projects used machine learning algorithms to predict future events or interpret collected data~\cite{Bishop-2006, Mitchell-1997, Hastie-2009, James-2013, Jordan-Mitchell-2015, Bird-2014, Bellinger-2017}.

\subsubsection{Prediction}
Prediction techniques aim to forecast the future accurately based on previous observations. Zheng {\em et al.} developed a framework to predict air quality readings of a monitoring station over the next 48 hours based on meteorological data, weather forecasts, and sensor readings from other nearby monitoring stations~\cite{Zheng-2015}. Azid {\em et al.} used principal component analysis and an artificial neural network to identify pollution sources and predict air pollution~\cite{Azid-2014}. Donnelly {\em et al.} combined kernel regression and multiple linear regression to forecast the concentrations of nitrogen dioxide over the next 24 and 48 hours~\cite{Donnelly-2015}. Hsieh {\em et al.} utilized a graphical model to predict the air quality of a given location grid based on data from sparse monitoring stations~\cite{Hsieh-2015}. These studies applied prediction techniques to help citizens plan daily activities and also inform regulators in controlling air pollution sources. Most of these studies focus on forecasting or interpolating sensing data. To the best of our knowledge, none of them considered human-reported data in their predictive models.

\subsubsection{Interpretation}
Interpretation techniques aim to extract knowledge from the collected data. This knowledge can help to discover potential interrelationships between predictors and responses, which is known to be essential in analyzing the impacts of environmental issues in the long-term~\cite{Brown-1992, Broeder-2016}. Gass {\em et al.} investigated the joint effects of outdoor air pollutants on emergency department visits for pediatric asthma by applying Decision Tree learning~\cite{Gass-2014}. The authors suggested using Decision Tree learning to hypothesize about potential joint effects of predictors for further investigation. Stingone {\em et al.} trained decision trees to identify possible interaction patterns between air pollutants and math test scores of kindergarten children~\cite{Stingone-2017}. Hochachka {\em et al.} fused traditional statistical techniques with boosted regression trees to extract species distribution patterns from the data collected via the eBird platform~\cite{Hochachka-2012}. These previous studies utilized domain knowledge to fit machine learning models with high explanatory powers on filtered citizen science data. In this paper, we also used Decision Tree to explore hidden interrelationships in the data. This extracted knowledge can reveal local concerns and serve as convincing evidence for communities in taking action.
	
	\section{Design Principles and Challenges}

Our goals are {\em (i)} to develop a system that can lower the barriers to contribute smell data and {\em (ii)} to make sure the data are useful in studying the impact of urban air pollution and advocating for better air quality. Each goal yields a set of design challenges.

\subsection{Collecting Smell Data at Scale with Ease}

Outside the scope of citizen science, a few works have collected human-reported smell data in various manners. However, these manners are not suitable for our projects. For example, prior works have applied a \textit{smell-walking} approach to record and map the landscape of smell experiences by recruiting participants to travel in cities~\cite{Henshaw-2013, Quercia-2015, Quercia-2016}. This process is labor intensive and hard for long-term air quality monitoring. Hsu {\em et al.} has also demonstrated that resident-reported smell reports, collected via Google Forms, can form evidence about air pollution when combined with data from cameras and air quality sensors~\cite{Hsu-2017, Hsu-2016}. While Google Forms is usable for a small-size study, it would not be effective in collecting smell reports on a city-wide scale with more than 300,000 affected residents over several years. Therefore, we developed a mobile system to record GPS locations and timestamps automatically. The system is specialized for gathering smell data at a broad temporal and geographical scale.

\subsection{What is Useful Data? A Wicked Problem}

There is a lack of research in understanding the potential of using smell as an indicator of urban air pollution. Moreover, we recognized that there are various methods of collecting, presenting, and using the data. It is not feasible to explore and evaluate all possible methods without deploying the system in the real-world context. These challenges constitute a \textit{wicked problem}~\cite{Conklin-2005, Rittel-1973}, which refers to problems that have no precise definition, cannot be fully observed at the beginning, are unique and depend on context, have no opportunities for trial and error, and have no optimal or \quo{right} solutions. In response to this challenge, our design principle is inspired by how architects and urban designers address wicked problems. When approaching a community or city-scale problem, architects and urban planners first explore problem attributes (as defined in~\cite{Pena-2012}) and then design specific solutions based on prior empirical experiences. We made use of an existing network of community advocacy groups, including ACCAN~\cite{ACCAN}, GASP~\cite{GASP}, Clean Air Council~\cite{CAC}, PennFuture~\cite{PennFuture}, and PennEnvironment~\cite{PennEnvironment}. These groups were pivotal in shaping the design of Smell Pittsburgh and providing insights into how to engage the broader Pittsburgh community. Through a series of informal community meetings, we iteratively co-designed the smell ratings (severity of odor), the description of each rating, and the wording of the questions (about smell source and symptom) with affected residents and activists in local advocacy groups.

Moreover, to sustain participation, we visualized smell report data on a map and also engage residents through push notifications. To add more weight to citizen-contributed pollution odor report, we engineered the application to send smell reports directly to the Allegheny County Health Department (ACHD). This strategy ensured that the local health department could access high-resolution citizen-generated pollution data to ascertain better and address potential pollution sources in our region. We met and worked with staff in ACHD to determine how they hoped to utilize smell report data and adjusted elements of the application to better suit their needs, such as sending data directly to their database and using these data as evidence of air pollution. Based on their feedback, the system submitted all smell reports to the health department, regardless of the smell rating. This approach provided ACHD with a more comprehensive picture of the local pollution landscape.

In summary, when developing Smell Pittsburgh, we considered the system as an ongoing infrastructure to sustain communities over time (as mentioned in~\cite{Dantec-2013}), rather than a software product which solves a single well-defined problem. The system is designed to influence citizen participation and reveals community concerns simultaneously, which is different from observational studies that use existing data, such as correlating air quality keywords from social media contents with environmental sensor measurements~\cite{Ford-2017}.
	
	\section{System}

Smell Pittsburgh is a system, distributed through iOS and Android devices, to collect smell reports and track urban pollution odors. We now describe two system features: (1) a mobile interface for submitting and visualizing odor complaints and (2) push notifications for predicting the potential presence of odor events.

\subsection{Submitting and Visualizing Smell Reports}

Users could report odor complaints via Smell Pittsburgh from their mobile devices via the submission console (Figure \ref{fig:UI_smell_pittsburgh}, left). To submit a report, users first selected a smell rating from 1 to 5, with one being \quo{just fine} and five being \quo{about as bad as it gets.} These ratings, their color, and the corresponding descriptions were designed to mimic the US EPA Air Quality Index \cite{EPA-2014}. Also, users could fill out optional text fields where they could describe the smell (e.g., industrial, rotten egg), their symptoms related to the odor (e.g., headache, irritation), and their personal experiences. Once a user submitted a smell report, the system sent it to the local health department and anonymously archived it on our backend database. Users could decide if they were willing to provide their contact information to the health department through the system setting panel. Regardless of the setting, our database did not record personal information.

The system visualized smell reports on a map that also depicted fine particulate matter and wind data from government-operated air quality monitoring stations (Figure \ref{fig:UI_smell_pittsburgh}, right). All smell reports were anonymous, and their geographical locations were skewed to preserve privacy. When clicking or tapping on the playback button, the application animated 24 hours of data for the currently selected day, which served as convincing evidence of air quality concerns. Triangular icons indicated smell reports with colors that correspond to smell ratings. Users could click on a triangle to view details of the associated report. Circular icons showed government-operated air quality sensor readings with colors based on the Air Quality Index \cite{EPA-2014} to indicate the severity of particulate pollution. Blue arrows showed wind directions measured from nearby monitoring stations. The timeline on the bottom of the map represented the concentration of smell reports per day with grayscale squares. Users could view data for a date by selecting the corresponding square.

\subsection{Sending Push Notifications}

Smell Pittsburgh sent post hoc and predictive event notifications to encourage participation. When there were a sufficient number of poor odor reports during the previous hour, the system sent a post hoc event notification: \quo{\highlightI{Many residents are reporting poor odors in Pittsburgh. Were you affected by this smell event? Be sure to submit a smell report!}} The intention of sending this notification was to encourage users to check and report if they had similar odor experiences. Second, to predict the occurrence of abnormal odors in the future, we applied machine learning to model the relationships between aggregated smell reports and air quality measurements from the past. We defined the timely and geographically aggregated reports as smell events, which indicated that there would be many high-rating smell reports within the next 8 hours. Each day, whenever the model predicted a smell event, the system sent a predictive notification: \quo{\highlightI{Local weather and pollution data indicates there may be a Pittsburgh smell event in the next few hours. Keep a nose out and report smells you notice.}} The goal of making the prediction was to support users in planning daily activities and encourage community members to pay attention to the air quality. To keep the prediction system updated, we computed a new machine learning model every Sunday night based on the data collected previously.
	
	\section{Evaluation}

The evaluation shows that using smell experiences is practical for revealing urban air quality concerns and empowering communities to advocate for a sustainable environment. Our goal is to evaluate the \highlightB{impact} of deploying interactive systems on communities rather than the usability (e.g., the time of completing tasks). We believe that it is more beneficial to ask \highlightB{\quo{Is the system influential?}} instead of \quo{Is the system useful?} We now discuss four studies: {\em (i)} system usage information of smell reports and interaction events, {\em (ii)} a dataset for predicting and interpreting smell event patterns, {\em (iii)} analysis of the effect of push notifications, and {\em (iv)} a survey of attitude changes and motivation factors.

\subsection{System Usage Study}

\begin{figure}[p]
	\centering
	\includegraphics[width=0.65\columnwidth]{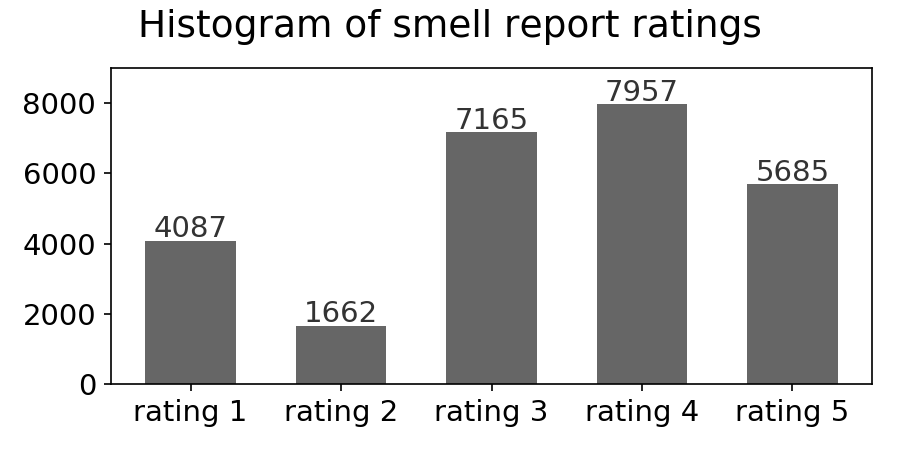}
	\caption{The distribution of the ratings (severity of the odor) in the smell reports. Our users tend to report bad odors with ratings greater than two.}
	\label{fig:all_smell_hist}
	\vspace{7mm}
	\centering
	\includegraphics[width=0.85\columnwidth]{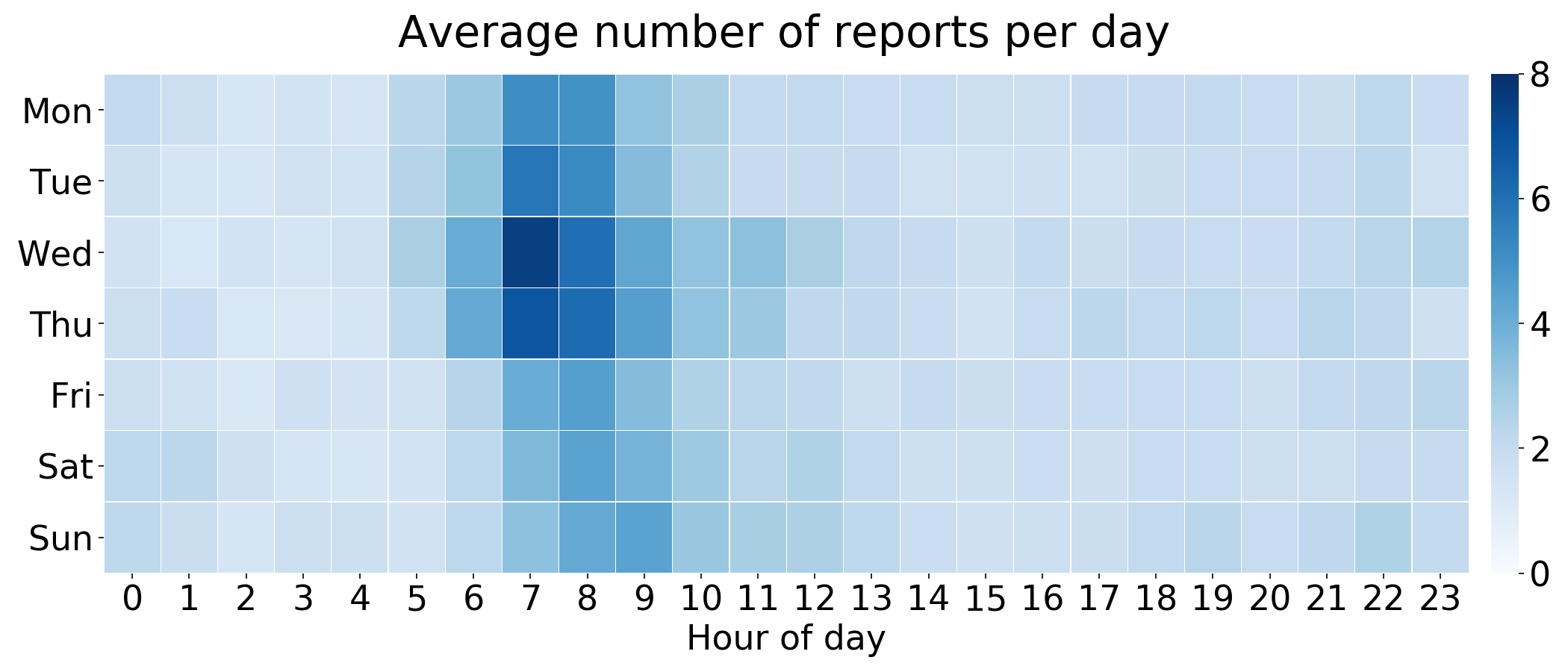}
	\caption{The average number of smell reports per day, aggregated by hour of day and day of week. Users usually submit reports in the morning and rarely at nighttime. Weekdays have more smell reports than weekends.}
	\label{fig:smell_day_hour}
	\vspace{7mm}
	\centering
	\includegraphics[width=0.85\columnwidth]{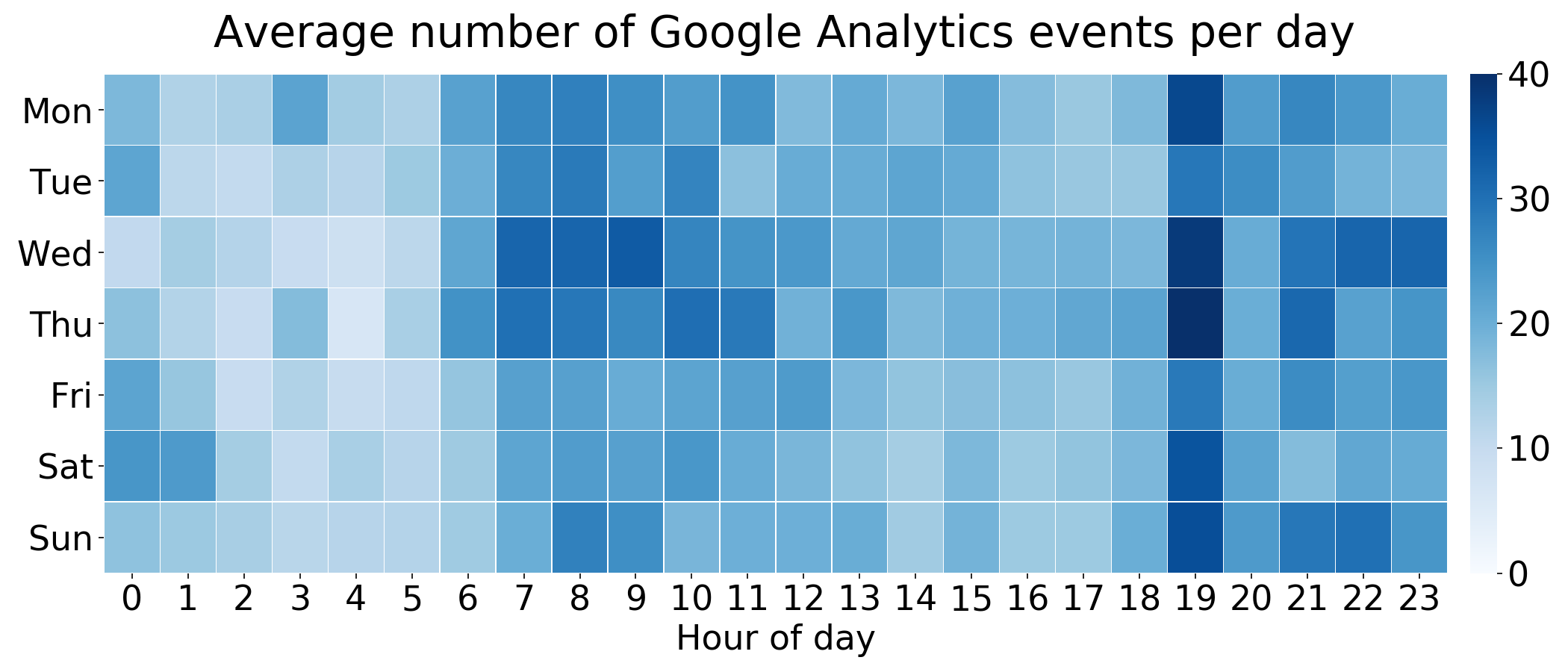}
	\caption{The average number of Google Analytics events per day, aggregated by hour of day and day of week. The dark vertical column at 7 pm is the time that the system distributes a summary notification to inform our users about how many reports were submitted for that day.}
	\label{fig:ga_day_hour}
\end{figure}

In this study, we show the usage patterns on mobile devices by parsing server logs and Google Analytics events. From our soft launch in November 2016 to the end of June 2019 over 32 months, there were 6,670 unique anonymous users (estimated by Google Analytics) in Pittsburgh. About 60\% and 40\% of the users use iOS and Android operation systems respectively. Our users contributed 26,556 smell reports, 862,943 alphanumeric characters in the submitted text fields, and 305,263 events of interacting with the visualization (e.g., clicking on icons on the map). About 78\% of the smell reports had ratings greater than two, showing that users tend to report bad odors (Figure~\ref{fig:all_smell_hist}).

\begin{table}[t]
	\caption{Percentages (rounded to the first decimal place) and the total size of different user groups. Abbreviation \quo{GA} means Google Analytics. Characters mean the number of characters that user entered in the text fields of smell reports.}
	\label{tab:user_group}
	\begin{tabular}{lcccc}
		\toprule
		& \multicolumn{4}{c}{Number of} \\ 
		& unique users & smell reports & characters & GA events \\
		\midrule
		Enthusiasts\!\!\! & 9.1\% & 55.5\% & 62.8\% & 47.3\% \\
		Explorers\!\!\! & 36.3\% & 37.3\% & 31.6\% & 27.3\% \\
		Contributors\!\!\! & 9.9\% & 7.2\% & 5.7\% & ... \\
		Observers\!\!\! & 44.8\% & ... & ... & 25.4\% \\
		\midrule
		Size (N)\!\!\! & 6,670 & 26,556 & 862,943 & 305,263 \\
		\bottomrule
	\end{tabular}
	\vspace{10mm}
	\caption{Statistics of user groups (median $\pm$ semi-interquartile range), rounded to the nearest integer. Symbol $\forall$ means \quo{for each.} Abbreviation \quo{GA} means Google Analytics. Characters mean the number of characters that user entered in the text fields of smell reports. Hours difference means the number of hours between the hit and data timestamps. A large difference between these timestamps means that users view the data (e.g., a smell report) that are far away from the current time.}
	\label{tab:user_group_vars}
	\begin{tabular}{lcccc}
		\toprule
		& \multicolumn{4}{c}{Number of} \\ 
		& \begin{tabular}{@{}c@{}}smell reports \\ $\forall$ user \end{tabular} \!\!\!
		& \begin{tabular}{@{}c@{}}characters \\ $\forall$ report \end{tabular} \!\!\!
		& \begin{tabular}{@{}c@{}}GA events \\ $\forall$ user \end{tabular} \!\!\!
		& \begin{tabular}{@{}c@{}}hours difference \\ $\forall$  event \end{tabular} \!\!\! \\
		\midrule
		Enthusiasts\!\!\! & 16$\pm$9 & 16$\pm$19 & 128$\pm$99 & 9$\pm$16 \\
		Explorers\!\!\! & 2$\pm$2 & 10$\pm$15 & 15$\pm$15 & 11$\pm$27 \\
		Contributors\!\!\! & 1$\pm$1 & 10$\pm$15 & ...  & ... \\
		Observers\!\!\! &  ... &  ...  &  9$\pm$9 & 19$\pm$49 \\
		\midrule
		All\!\!\! & 3$\pm$3 & 14$\pm$18 & 13$\pm$17 & 11$\pm$27 \\
		\bottomrule
	\end{tabular}
\end{table}

To investigate the distribution of smell reports and interaction events among our users, we divided all users into four types: enthusiasts, explorers, contributors, and observers (Table~\ref{tab:user_group}). Contributors submitted reports but did not interact with the visualization. Observers interacted with the visualization but did not submit reports. Enthusiasts submitted more than 6 reports and interacted with the visualization more than 30 times. Thresholds 6 and 30 were the median of the number of submitted reports and interaction events for all users respectively, plus their semi-interquartile ranges. Explorers submitted 1 to 6 reports or interacted with the visualization 1 to 30 times. We were interested in four variables with different distributions among user groups, which represented their characteristics (Table~\ref{tab:user_group_vars}). First, for each user, we computed the number of submitted reports and interaction events. Then, for each smell report, we calculated the number of alphanumeric characters in the submitted text fields. Finally, for interaction events that involved viewing previous data, we computed the time difference between hit timestamps and data timestamps. These two timestamps represented when users interacted with the system and when the data were archived respectively. Distributions of all variables differed from normal distributions (p\textless.001 for both D'Agostino's K-squared test and Shapiro-Wilk test) and were skewed toward zero.

The user group study showed highly skewed user contributions. About 34\% of the users submitted only one report. About 50\% and 81\% of the users submitted less than three and ten reports respectively, which aligned with the typical pattern in citizen science projects that many volunteers participated for only a few times~\cite{Sauermann-2015}. Moreover, these three user groups differed regarding the type and amount of data they contributed. Table~\ref{tab:user_group} shows that enthusiasts, corresponding to less than 10\% of the users, contributed about half of the data overall. Table~\ref{tab:user_group_vars} indicates the characteristics of these groups. Enthusiasts tended to contribute more smell reports, the number of alphanumeric characters of reports, and interaction events. Observers tended to browse data that were far away from the interaction time. Further investigation of the enthusiast group revealed a weak positive association (Pearson correlation coefficient r=.32, n=604, p\textless.001) between the number of submitted reports and the number of user interaction events.

\begin{figure}[t]
	\centering
	\includegraphics[width=0.7\columnwidth]{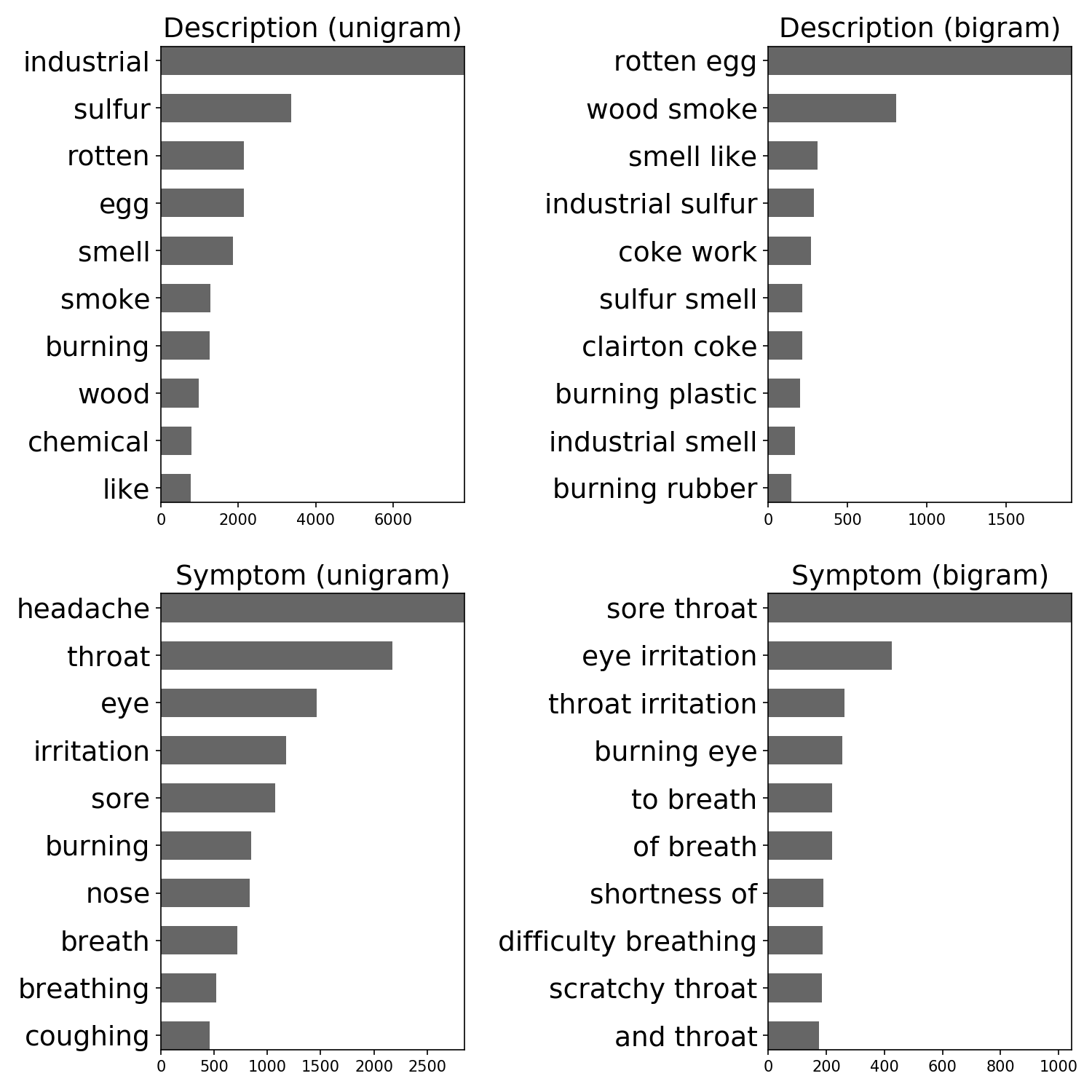}
	\caption{Text analysis of high frequency words (unigram) and phrases (bigram) in the text fields of all smell reports. Most of them describe industrial pollution odors and symptoms of air pollution exposure, especially hydrogen sulfide (rotten egg smell).}
	\label{fig:text_analysis}
\end{figure}

To identify critical topics in citizen-contributed smell reports, we analyzed the frequency of words (unigram) and phrases (bigram) in the text fields. We used python NLTK package~\cite{Bird-2009} to remove stop words and group similar words with different forms (lemmatization). Figure~\ref{fig:text_analysis} shows that high-frequency words and phrases mostly described industrial pollution odors and symptoms of air pollution exposure, especially hydrogen sulfide that has rotten egg smell and can cause a headache, dizziness, eye irritation, sore throat, cough, nausea, and shortness of breath~\cite{Lindenmann-2010, Reiffenstein-1992, Guidotti-2010, NRC-2009}. This finding inspired us to examine how hydrogen sulfide affected urban odors in the next study.

\subsection{Smell Dataset Study}

\begin{figure}[t]
	\centering
	\includegraphics[width=1\columnwidth]{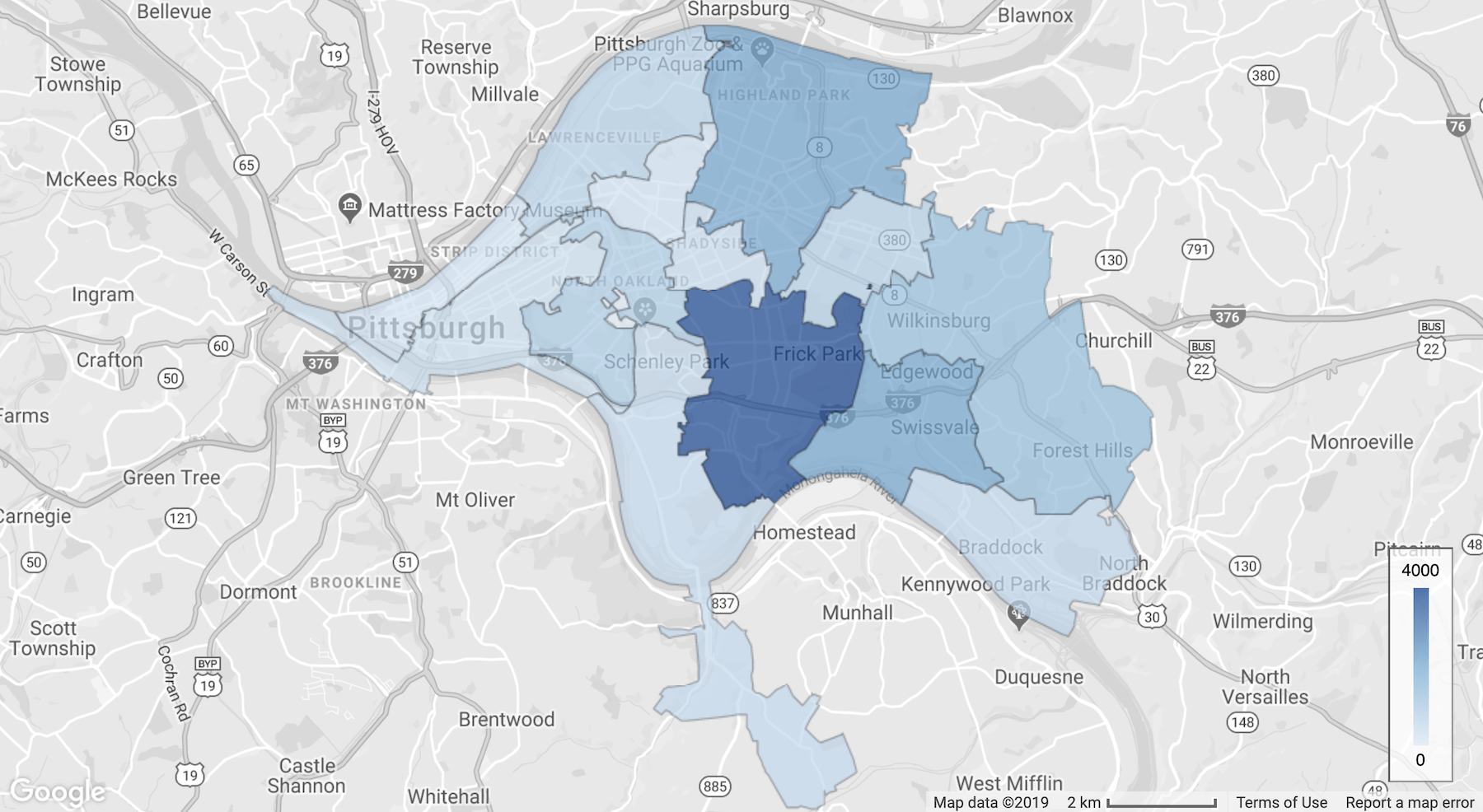}
	\caption{The distribution of the number of smell reports on selected zip code regions. Note that the numbers reflect the level of user engagement but do not necessarily represent the severity of the odor pollution.}
	\label{fig:smell-geo}
\end{figure}

In this study, we show that human-reported smell data, despite noisy, can still enable prediction and contribute scientific knowledge of interpretable air pollution patterns. We first constructed a dataset with air quality sensor readings and smell reports from October 31 in 2016 to September 27 in 2018~\cite{smell-dataset-2018}. The sensor data were recorded hourly by twelve government-operated monitoring stations at different locations in Pittsburgh, which included timestamps, particulate matters, sulfur dioxide, carbon monoxide, nitrogen oxides, ozone, hydrogen sulfide, and wind information (direction, speed, and standard deviation of direction). The smell report data contained timestamps, zip-codes, smell ratings, descriptions of sources, symptoms, and comments. For privacy preservation, we dropped the GPS location (latitude and longitude) of the smell reports and used zip-codes instead.

We framed the smell event prediction as a supervised learning task to approximate the function $F$ that maps a predictor matrix $X$ to a response vector $y$. The predictor matrix and the response vector represented air quality data and smell events respectively. To build $X$, we re-sampled air quality data over the previous hour at the beginning of each hour. For example, at 3 pm, we took the mean value of sensor readings between 2 pm and 3 pm to construct a new sample. Wind directions were further decomposed into cosine and sine components. To equalize the effect of predictors, we normalized each column of matrix $X$ to zero mean and unit variance. Missing values were replaced with the corresponding mean values.

To build $y$ that represents smell events, we aggregated high-rating smell reports over the future 8 hours at the beginning of each hour. We specifically chose the geographic regions that have sufficient amount of data during aggregation (Figure~\ref{fig:smell-geo}). For instance, at 6 am, we took the sum of smell ratings with values higher than two between 6 am and 2 pm to obtain a confidence score, which represented agreements of how likely a smell event occurred. The scores were further divided into positive and negative classes (with or without smell events) by using threshold 40. In this way, we simplify the task to a binary classification problem, with 64 predictor variables (columns of $X$) and 16,766 samples (rows of $X$ and $y$). The distribution of classes was highly imbalanced (only 8\% positive). Besides classification, we also applied a regression approach to predict the confidence scores directly without thresholding initially. Then the predicted scores were thresholded post hoc with value 40 to produce positive and negative classes, which enabled us to compare the performance of these two approaches.

When performing classification and regression, we added 3-hour lagged predictor variables, days of the week, hours of the day, and days of the month into the original predictor variable, which expanded its length from 64 to 195. We chose the lagged duration during model selection. We implemented two models, Random Forest~\cite{Breiman-2001} and Extremely Randomized Trees~\cite{Geurts-2006}, by using python scikit-learn package~\cite{Pedregosa-2011}. These algorithms build a collection of decision trees using the CART algorithm~\cite{Breiman-1984}, where the leaves represent the responses $y$ and the branches represent the logical conjunction of predictors in $X$. There were three tunable model parameters: the number of trees, the number of features to select randomly for splitting a tree node, and the minimum number of samples required to split a tree node. For simplicity, we fixed the number of trees (1,000 for classification and 200 for regression) and chose other parameters during model selection.

\begin{figure}[t]
	\centering
	\includegraphics[width=0.7\columnwidth]{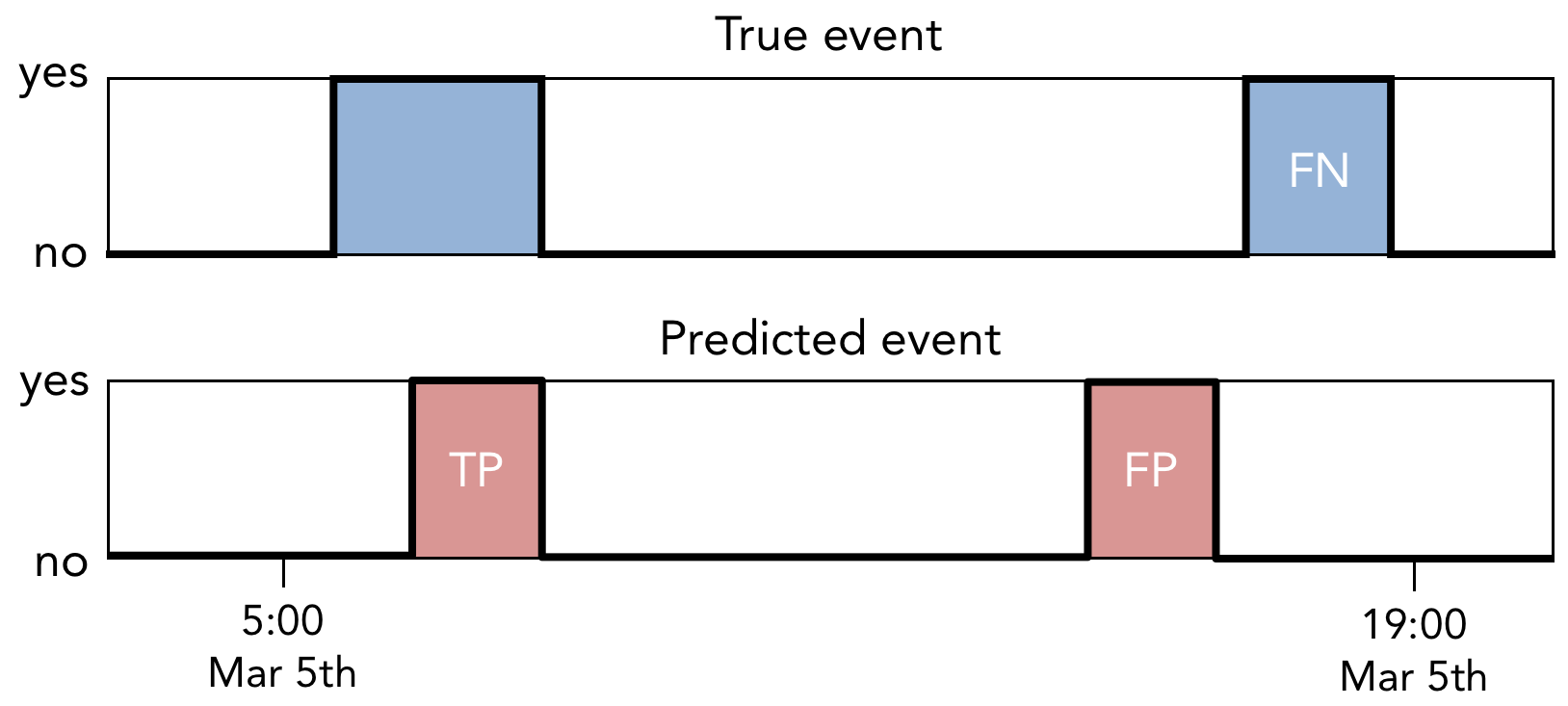}
	\caption{This figure shows true positives (TP), false positives (FP), and false negatives (FN). The x-axis represents time. The blue and red boxes indicate ground truth and predicted smell events respectively.}
	\label{fig:metrics}
\end{figure}

To evaluate model performance, we defined and computed true positives (TP), false positives (FP), and false negatives (FN) to obtain precision, recall, and F-score \cite{Powers-2011} (Figure \ref{fig:metrics}). We first merged consecutive positive samples to compute the starting and ending time of smell events. Then, if a predicted event overlapped with a ground truth event, we counted this event as a TP. Otherwise, we counted a non-overlapped predicted event as an FP. For ground truth events that had no overlapping predicted events, we counted them as FN. When computing these metrics, we considered only daytime events because residents rarely submitted reports during nighttime (Figure \ref{fig:smell_day_hour}). We defined daytime from 5 am to 7 pm. Since the model predicted if a smell event would occur in the next 8 hours, we only evaluated the prediction generated from 5 am to 11 am.

\begin{table}[t]
	\caption{Cross-validation of model performances (mean $\pm$ standard deviation) on the testing set for daytime. We run this experiment for 100 times with random seeds. Abbreviations \quo{ET} and \quo{RF} indicate Extremely Randomized Trees and Random Forest respectively, which are used for predicting upcoming smell events.}
	\label{tab:prediction}
	\begin{tabular}{lccc}
		\toprule
		& Precision & Recall & F-score \\
		\midrule
		Classification ET\!\!\!
		& 0.87$\pm$0.01 & 0.59$\pm$0.01 & 0.70$\pm$0.01 \\
		Classification RF\!\!\!
		& 0.80$\pm$0.02 & 0.57$\pm$0.01 & 0.66$\pm$0.01 \\
		Regression ET\!\!\!
		& 0.57$\pm$0.01 & 0.76$\pm$0.01 & 0.65$\pm$0.01 \\
		Regression RF\!\!\!
		& 0.54$\pm$0.01 & 0.75$\pm$0.01 & 0.63$\pm$0.01 \\
		\bottomrule
	\end{tabular}
\end{table}

We chose model parameters by using time-series cross-validation \cite{Kohavi-1995, Arlot-2010}, where the dataset was partitioned and rolled into several pairs of training and testing subsets for evaluation. Because our predictors and responses were time-dependent, we used previous samples to train the models and evaluated them on future data. We first divided all samples into folds, with each fold approximately representing a week. Then, starting from fold 49, we took the previous 48 folds as training data (about 8,000 samples) and the current fold as testing data (about 168 samples). We iterated this procedure for the rest of the folds, which reflected the setting of the deployed system where a new model was trained on every Sunday night by using data from the previous 48 weeks. Table \ref{tab:prediction} reports the evaluation metrics after cross-validating the models 100 times with various random seeds.

While these models enabled us to predict future smell events, they were typically considered as black box models and not suitable for interpreting patterns. Although these models provided feature importances, interpreting these weights could be problematic because several predictors in the dataset were highly correlated, which might appear less significant than other uncorrelated counterparts. Inspired by several previous works related to extracting knowledge from data \cite{Shaikhina-2017, Gass-2014, Caruana-2006}, we utilized a white box model, Decision Tree, to explain a representative subset of predictors and samples, which were selected by applying feature selection \cite{Guyon-2003} and cluster analysis. One can view this process as performing model compression to distill the knowledge in a large black box model into a compact model that is explainable to human \cite{Bucilua-2006, Hinton-2015}.

\begin{figure*}[t]
	\centering
	\includegraphics[width=1\columnwidth]{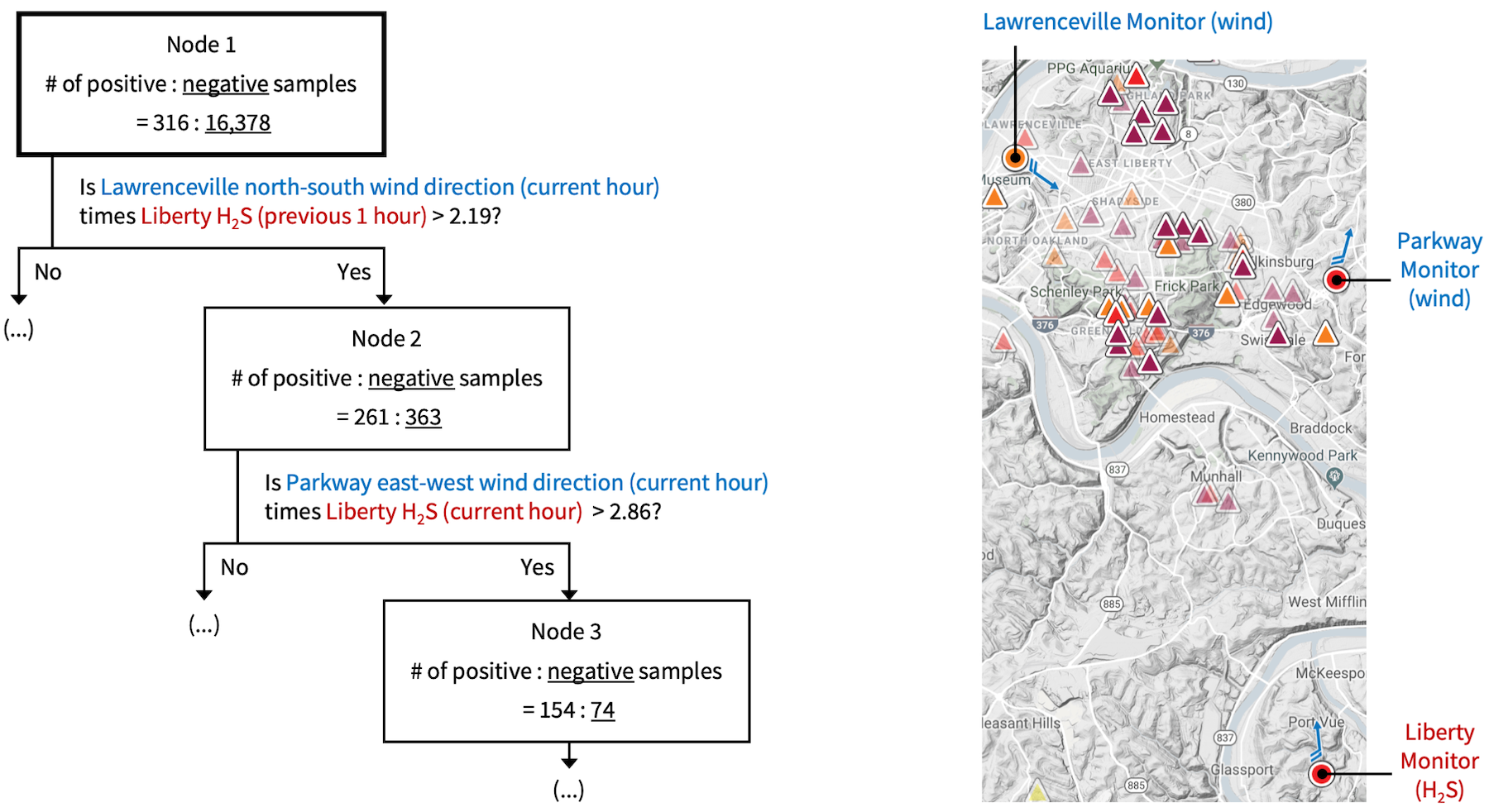}
	\caption{The right map shows smell reports and sensor readings at 10:30 am on December 3, 2017. The left graph shows one example of the Decision Tree, which explains the pattern of about 30\% smell events. Each tree node indicates the ratio of the number of positive (with smell event) and negative samples (no smell event). The most important feature is the interaction between the current north-south wind directions at Lawrenceville and the previous 1-hour hydrogen sulfide readings at Liberty ($r$=.58, $p$\textless.001, $n$=16,694), where $p$ means the p-value, $n$ means the number of samples, and $r$ means the point-biserial correlation of the predictor and smell events. The second-most important feature is the interaction between the current east-west wind directions at Parkway and the current hydrogen sulfide readings at Liberty ($r$=.54, $p$\textless.001, $n$=16,694). The corresponding Gini importance for the most and the second-most important features are 0.42$\pm$0.02 and 0.10$\pm$0.00, respectively, in the format of \quo{mean $\pm$ standard deviation} for the 100-times experiment.}
	\label{fig:dt}
\end{figure*}

\begin{table}[t]
	\caption{Cross-validation of model performances (mean $\pm$ standard deviation) on the training and testing set for daytime. We run this experiment for 100 times, including both the cluster analysis and recursive feature elimination. The Decision Tree is used for interpreting air pollution patterns on a subset of the entire dataset.}
	\label{tab:interpretation}
	\begin{tabular}{lcccc}
		\toprule
		& Precision & Recall & F-score & Phase \\
		\midrule
		Decision Tree\!\!\!
		& 0.79$\pm$0.00 & 0.86$\pm$0.03 & 0.82$\pm$0.01 & training \\
		Decision Tree\!\!\!
		& 0.49$\pm$0.01 & 0.62$\pm$0.04 & 0.54$\pm$0.03 & testing \\
		\bottomrule
	\end{tabular}
\end{table}

During data interpretation, we only considered the classification approach due to better performance. First, we used domain knowledge to manually select features. As there were many highly correlated features, selecting a subset of them arbitrarily for extracting patterns was impractical. The knowledge obtained from informal community meetings and the result discovered in the text analysis (Figure \ref{fig:text_analysis}) suggested that hydrogen sulfide might be the primary source of smell events. This finding inspired us to chose hydrogen sulfide, wind direction, wind speed, and standard deviation of wind direction from all of the other available predictors. The current and up to 2-hour lagged predictor variables were all included. Also, we added interaction terms of all predictors, such as hydrogen sulfide multiplied by the sine component of wind direction. This manual feature selection procedure produced 781 features.

Then, we used DBSCAN \cite{Ester-1996} to cluster positive samples and to choose a representative subset. The distance matrix for clustering was derived from a Random Forest in an unsupervised manner, as described in~\cite{Shi-2006-unsupervised}. The goal was to construct another synthetic dataset and use both datasets to predict if a data sample came from the original one. For each predictor in the original dataset, we sampled the predictor values for $n$ times uniformly at random (with replacement) and put the values in the synthetic dataset for the corresponding predictor, where $n$ is the size of the original dataset. After fitting the model, for each sample pair, we counted the number of times that the pair appeared in the same leaf for all trees in the model. The numbers were treated as the similarity of sample pairs, and we scaled the similarity by dividing it with the number of trees in the model. We converted the similarity $s$ into distance $d$ by using $d=1-s$. This procedure identified a cluster with about 25\% of positive samples.

Finally, we used recursive feature elimination \cite{Guyon-2002} to remove 50 features that had the smallest weights iteratively, which resulted in 30 most important features. These feature importance weights were computed by fitting a Random Forest. We trained a Decision Tree using the CART algorithm \cite{Breiman-1984} to interpret the cluster and the selected 30 features. Parameters for data interpretation (DBSCAN, Random Forest, and Decision Tree) were selected by using cross-validation. Table \ref{tab:interpretation} reported the evaluation metrics after running the data interpretation pipeline for 100 times, including the cluster analysis and recursive feature elimination. The result showed that the model explains the underlying pattern of about 30\% of the smell events, which was a joint effect of wind information and hydrogen sulfide readings. Figure \ref{fig:dt} indicated the model and the most important two features, which were consistent for 99 times, out of the total 100 times experiment.

\subsection{Push Notifications Study}\label{sec:pn_study}

Through randomized experiments, using push notifications for mobile applications has been proven useful in engaging users. Bidargaddi {\em et al.} conducted a randomized experiment with 1,255 users of a well-being mobile app for 89 days and observed that users were more likely to engage with the app in the next 24 hours when a notification was sent~\cite{Niranjan-2018}. Phame {\em et al.} studied push notification uses of an English learning mobile app for a month and concluded that push notifications increase user engagement, but overly-frequent prompts result in an opposite effect~\cite{Pham-2016}. Freyne {\em et al.} also suggested that push notifications are appropriate mechanisms to engage users with mobile technology in the short term, while the persuasive power of prompts could vanish after a while~\cite{Freyne-2017}. Different from the prior works that applied randomized experiments, this observational study investigates the effect of different types of push notifications on user engagement post hoc by parsing server logs and Google Analytics events. From November 2016 to June 2019 (32 months), the system has distributed 4,219 notifications. Among them, there are six types for various purposes:
\begin{itemize}
	\item \textbf{P1}: smell events that are predicted by the machine learning algorithm (discussed in the previous study) to indicate that Pittsburgh may suffer from poor odor in the next 8 hours
	\item \textbf{P2}: smell events that are observed post hoc to inform users that many others have submitted smell reports during the previous hour, which confirmed the poor odor problem
	\item \textbf{P3}: notices of the first submitted smell report for each day to notify people that someone has experienced poor odor
	\item \textbf{P4} and \textbf{P5}: notices of air quality changes to show if the air quality index dramatically improves or worsens when compared to the previous hour
	\item \textbf{P6}: a daily summary of the number of submitted smell reports to demonstrate the severe level of the poor odor problem
\end{itemize}
Table~\ref{tab:pn_type} contains examples of these notifications. We measure the effect within a time range by using six metrics: number of smell reports, unique users who submit the reports, unique zip codes in the reports, interaction events from Google Analytics, unique users who contribute the events, and pageviews. Notice that interaction events do not involve pageviews. The time range depends on the study type. We exclude all interactions from desktop devices and only focus on the behavior of mobile device users. Due to technical challenges, we did not include the number of unique zip codes in the interaction events as a metric.

\begin{table}[p]
	\caption{Summary of the push notification analysis. A checkmark (\boldcheckmark \checkmark \dashcheckmark) shows that the effect (measured by the metric) of the notification is statistically significant (p\textless.05), where \boldcheckmark and \dashcheckmark show that the regression model has identified a strong ($R^2$>.6 and $\hat{\beta}$>.6) or week ($R^2$<.4 and $\hat{\beta}$<.4) effect size respectively. Symbol \quo{...} shows an inconsistency between the rank test and the regression model, which means that the effect comes mainly from the confounders.}
	\label{tab:pn_type}
	\begin{tabular}{ll}
		\toprule 
		\multicolumn{2}{c}{\textbf{User engagement metrics}} \\ 
		Metric & Explanation\\
		\midrule
		\rowcolor{lightgray}
		M1 & Number of submitted smell reports \\
		M2 & Number of unique users who submit the reports \\
		\rowcolor{lightgray}
		M3 & Number of unique zipcodes in the smell reports \\
		M4 & Number of Google Analytics events \\
		\rowcolor{lightgray}
		M5 & Number of unique users who contribute the events \\
		M6 & Number of Google Analytics pageviews \\
		\bottomrule
		\vspace{3pt}
	\end{tabular}
	\begin{tabular}{clcccccc}
		\toprule
		\multicolumn{7}{c}{\textbf{Summary of the push notification analysis}} \\ 
		Type & Content & M1 & M2 & M3 & M4 & M5 & M6 \\
		\midrule
		\rowcolor{lightgray}
		P1 & \begin{tabular}[t]{@{}l@{}l@{}l@{}}
			\textbf{Smell Event Alert} \\
			Local weather and pollution data  indicates there \\
			may be a Pittsburgh smell event in the next few \\
			hours. Keep a nose out and report smells you notice!
		\end{tabular} & \boldcheckmark & \boldcheckmark & \boldcheckmark & \checkmark & \boldcheckmark & \boldcheckmark \vspace{3pt} \\
		P2 & \begin{tabular}[t]{@{}l@{}l@{}l@{}}
			\textbf{Smell Event Alert} \\
			Many residents are reporting poor odors in \\
			Pittsburgh. Were you affected by this smell event? \\
			Be sure to submit a smell report!
		\end{tabular} & & & & \checkmark & \checkmark & \checkmark \vspace{3pt} \\
		\rowcolor{lightgray}
		P3 & \begin{tabular}[t]{@{}l@{}}
			\textbf{How does your air smell?} \\
			A smell report rated 5 was just submitted.
		\end{tabular} & \boldcheckmark & \boldcheckmark & \boldcheckmark & \checkmark & \boldcheckmark & \boldcheckmark \vspace{3pt} \\
		P4 & \begin{tabular}[t]{@{}l@{}}
			\textbf{Does it smell better?} \\
			Pittsburgh AQI just improved.
		\end{tabular} & \dashcheckmark & \dashcheckmark & \dashcheckmark & & & ... \vspace{3pt} \\
		\rowcolor{lightgray}
		P5 & \begin{tabular}[t]{@{}l@{}l@{}}
			\textbf{PGH Air Quality Notification} \\
			PGH Air Quality Notification AQI \\
			has been over 50 for last 2 hrs.
		\end{tabular} & & & & ... & \dashcheckmark & \checkmark \vspace{3pt} \\
		P6 & \begin{tabular}[t]{@{}l@{}}
			\textbf{Smell Report Summary} \\
			53 smell reports were submitted today.
		\end{tabular} & \dashcheckmark & \dashcheckmark & \dashcheckmark & \checkmark & \checkmark & \checkmark \\
		\bottomrule
		\vspace{3pt}
	\end{tabular}
	\begin{tabular}{cccc}
		\toprule
		\multicolumn{4}{c}{\textbf{Time range and size of notifications}} \\ 
		Type & Size (N) & From & To \\
		\midrule
		\rowcolor{lightgray}
		P1 & 71 & 2018-01-22 & 2019-06-29 \\
		P2 & 79 & 2018-02-27 & 2019-06-29 \\
		\rowcolor{lightgray}
		P3 & 377 & 2016-11-14 & 2018-03-20 \\
		P4 & 753 & 2016-11-10 & 2019-06-30 \\
		\rowcolor{lightgray}
		P5 & 1,978 & 2016-11-10 & 2019-06-30 \\
		P6 & 961 & 2016-11-10 & 2019-06-30 \\
		\bottomrule
	\end{tabular}
\end{table}

We conduct two sub-studies in this analysis. In the first sub-study, we show whether user engagement changes in the short term after sending push notifications. For each type of notification, we compute the metrics within two hours before and after distributing it. For instance, if there was a notification at 10 am, we compute the metric (e.g., number of smell reports) from 8 am to 10 am and also the one from 10 am to 12 pm. In this way, we obtain two groups of paired measurements for each metric, which is the dependent variable $Y$. Sending the push notification can be viewed as a treatment, which is analogous to the treatment in medical studies. These two groups indicate the outcome with or without the treatment, which is the independent variable $X$. Since the dependent variables are not normally distributed (p\textless.05 for both D'Agostino's K-squared test and Shapiro-Wilk test), we apply the Wilcoxon signed-rank test to analyze if there exist statistically significant differences between the median values of these two groups. The rank test can be viewed as a nonparametric paired t-test. The null hypothesis of the rank test is that the median difference (instead of mean difference in paired t-test) between pairs of measurements is zero. A small p-value (\textless.05) in the rank test indicates that we favor the alternative hypothesis, which indicates shifts in the median values between two groups. Details of the rank test (sample size, mean, semi-interquartile range, p-value, and effect size) are in Table~\ref{tab:pn_analysis_pn1}, \ref{tab:pn_analysis_pn2}, \ref{tab:pn_analysis_pn3}, \ref{tab:pn_analysis_pn4}, \ref{tab:pn_analysis_pn5}, and \ref{tab:pn_analysis_pn6} in Appendix~\ref{ap:pn_study_other_tables}.

In the second sub-study, for the predictive notifications (type P1) generated by the machine learning model (discussed in the smell dataset study), we further analyze if they are effective across different days by comparing true positives and false negatives. A true positive (TP) means that there was a smell event, and the system successfully sent the algorithm-predicted notifications. A false negative (FN) means that there was a smell event, but the system failed to predict it. More precisely, in the case that the 8-hour prediction (e.g., from 6 am to 2 pm) indicated no smell event, we define this prediction a false negative if the sum of smell ratings with values higher than two in this 8-hour time range is larger than 40. No notifications were sent in these false negative cases. We exclude the false positive (FP) cases since the machine learning model is tuned to have low FP, and we do not have sufficient FP samples for analysis. For each TP and FN, We compute the metrics within two hours after their timestamps. For instance, if a TP happened at 8 am, we compute the metric (e.g., number of smell reports) from 8 am to 10 am. In this way, we obtain two groups (TP and FN) of non-paired measurements for each metric as the dependent variable $Y$. Since the dependent variable is not normally distributed (p\textless.05 for both D'Agostino's K-squared test and Shapiro-Wilk test) and not paired, we apply the Mann-Whitney U test, which can be viewed as the non-parametric version of the t-test. A small p-value (\textless.05) in the U test indicates a statistically significant difference in the median values between the two groups. Details of the Mann-Whitney U test (sample size, mean, semi-interquartile range, p-value, and effect size) are in Table~\ref{tab:pn_analysis_tp_fn}.

\begin{table}[t]
	\caption{Explanation of predictors (both the treatment and confounders) in the regression analysis. The particulate matters predictor (PM) aggregates PM2.5 and PM10. Abbreviation \quo{AQI} means air quality index. Predictors P1 to P6 (discussed in Table~\ref{tab:pn_type}) do not involve the push notification that we wish to study (the treatment). Predictors XH, YH, XM, and YM are the cyclical transformations for timestamps.}
	\label{tab:pn_predictors}
	\begin{tabular}{lll}
		\toprule
		Predictor & Data type & Explanation \\
		\midrule
		\rowcolor{lightgray} 
		treatment & binary & If the system sent the notification type that we wish to study \\
		P1 & binary & If type P1 notification occurred within the study time range \\
		\rowcolor{lightgray} 
		P2 & binary & If type P2 notification occurred within the study time range \\
		P3 & binary & If type P3 notification occurred within the study time range \\
		\rowcolor{lightgray} 
		P4 & binary & If type P4 notification occurred within the study time range \\
		P5 & binary & If type P5 notification occurred within the study time range \\
		\rowcolor{lightgray} 
		P6 & binary & If type P6 notification occurred within the study time range \\
		WD & binary & If the date of the treatment is weekday (Monday to Friday) \\
		\rowcolor{lightgray} 
		XH & interval & $\cos(\pi \cdot \mathrm{hour}/12)$, the cosine component for hour of day \\
		YH & interval & $\sin(\pi \cdot \mathrm{hour}/12)$, the sine component for hour of day\\
		\rowcolor{lightgray} 
		XM & interval & $\cos(\pi \cdot \mathrm{month}/6)$, the cosine component for month of year \\
		YM & interval & $\sin(\pi \cdot \mathrm{month}/6)$, the sine component for month of year \\
		\rowcolor{lightgray}
		E & interval & Elapsed time (in days) since the initial system soft launch date \\
		M & interval & The daily sum of the corresponding measured metric $Y$\\
		\rowcolor{lightgray} 
		O3 & interval & Daily AQI of ozone on the date of the treatment \\
		PM & interval & Daily AQI of particulate matters on the date of the treatment \\
		\rowcolor{lightgray} 
		CO & interval & Daily AQI of carbon monoxide on the date of the treatment \\
		SO2 & interval & Daily AQI of sulfur dioxide on the date of the treatment \\
		\bottomrule
	\end{tabular}
\end{table}

Furthermore, for all sub-studies, we conduct a regression analysis to determine if the effect truly comes from the treatment, instead of only the confounders. Table~\ref{tab:pn_predictors} explains these predictors in detail. Both the dependent (i.e., treatment) and independent (i.e., metrics) variables can be influenced by the confounders, which include daily air quality index of major pollutants, transformed features of time, elapsed time since the initial system soft launch date, the daily sum of the corresponding measured metric, and the existence of other notifications (excluding the treatment) within the measured time range. We compute the daily air quality index according to the EPA standard~\cite{EPA-2014}, and the air quality data were obtained from government-operated monitoring stations at different locations in Pittsburgh. Specifically, for the second sub-study, we extend the time range of the regression analysis to eight hours for every two hours after sending type P1 notification. For instance, if a true positive occurred at 8 am, we compute four groups of metrics: from 8 am to 10 am, from 10 am to 12 pm, from 12 pm to 2 pm, and from 2 pm to 4 pm. In this way, we can examine the effect of predictive notifications over time.

In the regression analysis, since our data violate the linearity and normality assumptions in multiple linear regression, we apply the Generalized Linear Model (GLM)~\cite{McCullagh-1989,Fox-2015,Faraway-2016,Dobson-2008} to predict user engagement metrics by using the treatment and the confounders. Assumptions of GLM include:
\begin{itemize}
	\item Each measured metric is independent and identically distributed.
	\item The metrics are affected by the transformation of a linear combination of the predictors.
	\item The distributions of the measured metrics follow some exponential family distributions.
\end{itemize}
In this way, we can examine the effect of the treatment on the measured metrics while adjusting for the confounders. Specifically, we aim to estimate the conditional mean $\epv(Y|X)$ of the model:
\begin{equation}\label{eq:full}
g\big(\epv(Y|X)\big) = \beta^T X = \beta_0 + \beta_1 X_1 + \beta_2 X_2 + ... + \beta_{m-1} X_{m-1}
\end{equation}
where $g$ is a link function, $\epv$ is the expected value, $\beta \in \real^{m \times 1}$ is the coefficient vector, $Y \in \real^{1 \times n}$ is the response vector, $X \in \real^{m \times n}$ is the predictor matrix, $m$ is the number of parameters (both predictors and intercept), and $n$ is the number of observations. In our case, $X$ includes the treatment and the confounders (Table~\ref{tab:pn_predictors}), $Y$ includes the user engagement metrics (Table~\ref{tab:pn_type}), $g$ is a natural log function, and $Y|X$ is assumed to follow a negative binomial distribution. For implementation, we use python and the statsmodels package~\cite{Seabold-2010} to fit the GLM with the Iteratively Reweighted Least Squares algorithm~\cite{Green-1984}. To reduce the number of predictors and simplify the model, we apply the elastic net regularization~\cite{Zou-2005}. For tuning the model and regularization parameters, we perform a grid search and choose the model with the lowest Akaike Information Criterion score~\cite{Akaike-1974}. To check if some predictors are highly related to a linear combination of the others, we compute the Variance Inflation Factors (VIF) and find that our data do not suffer from such multicollinearity. To select the models that fit the data well, we conduct the goodness of fit test using deviance and Pearson statistics. We also apply a zero-coefficient test with the Wald statistic to identify the effect of the treatment and confounders. A small p-value (\textless.05) indicates that the corresponding type of push notification is effective. An overview of GLM and the implementation details are in Appendix~\ref{ap:glm}.

\begin{figure}[p]
	\centering
	\includegraphics[width=1\columnwidth]{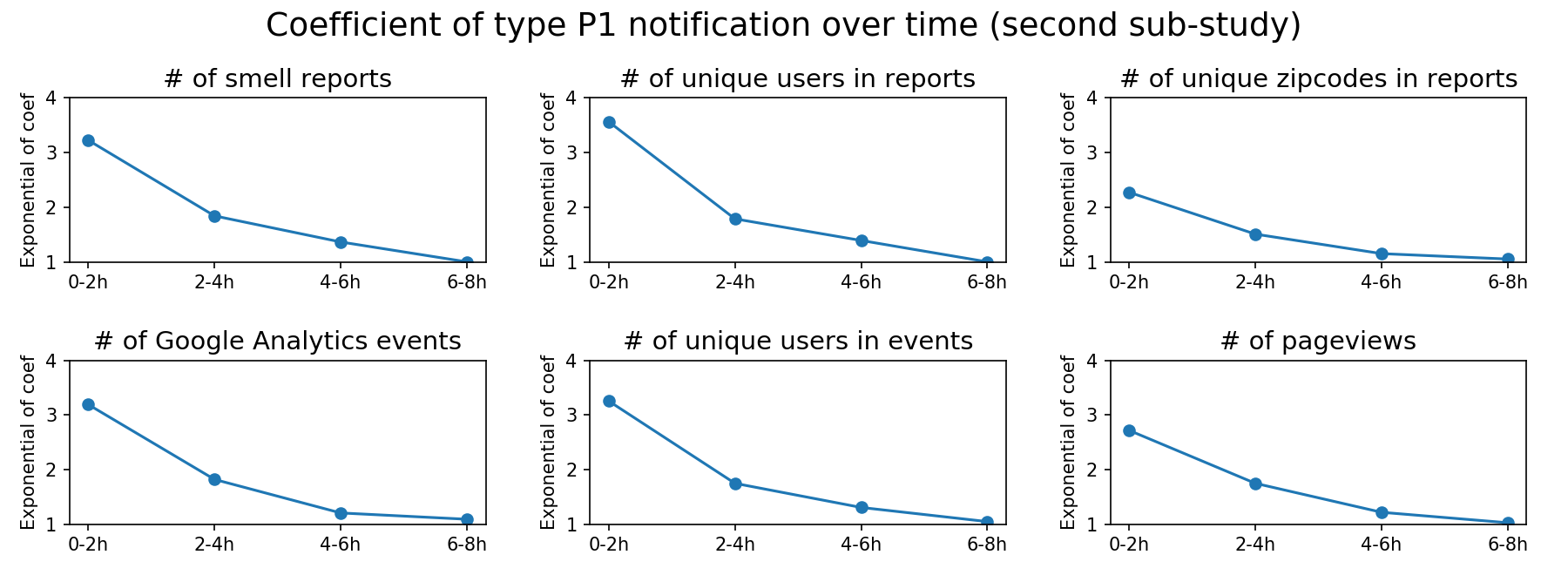}
	\vspace{2mm} \ \\
	\includegraphics[width=1\columnwidth]{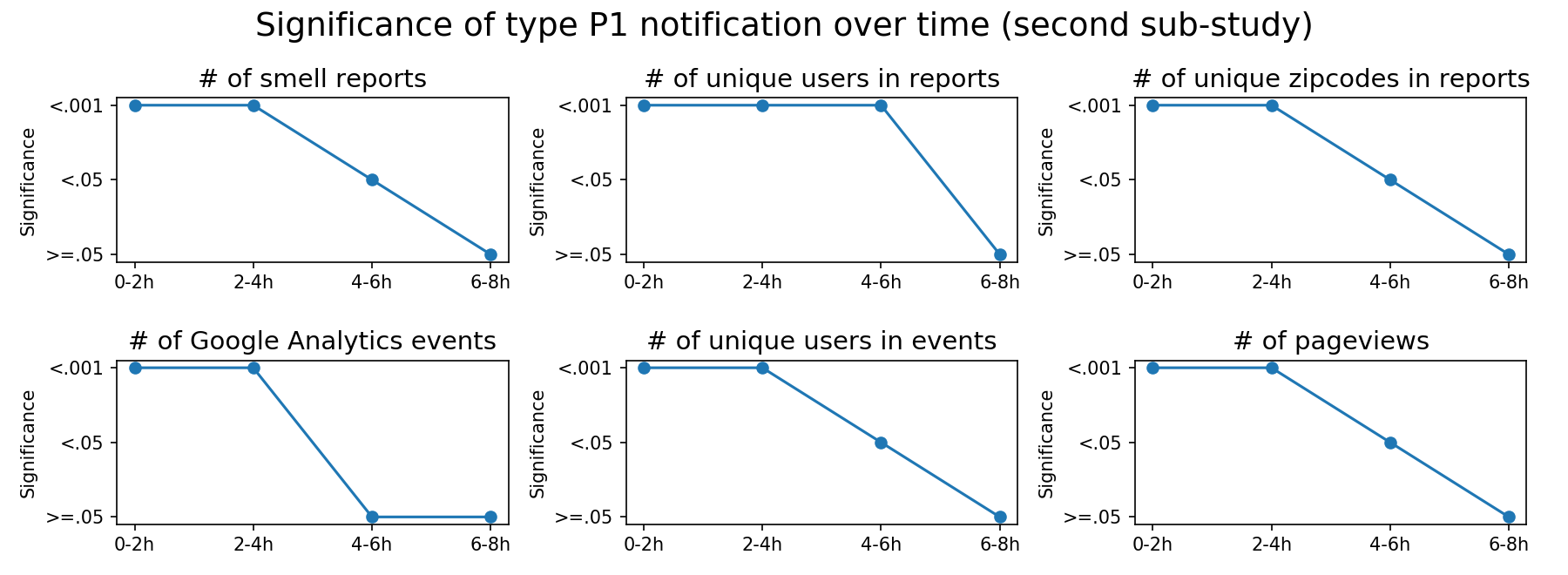}
	\vspace{2mm} \ \\
	\includegraphics[width=1\columnwidth]{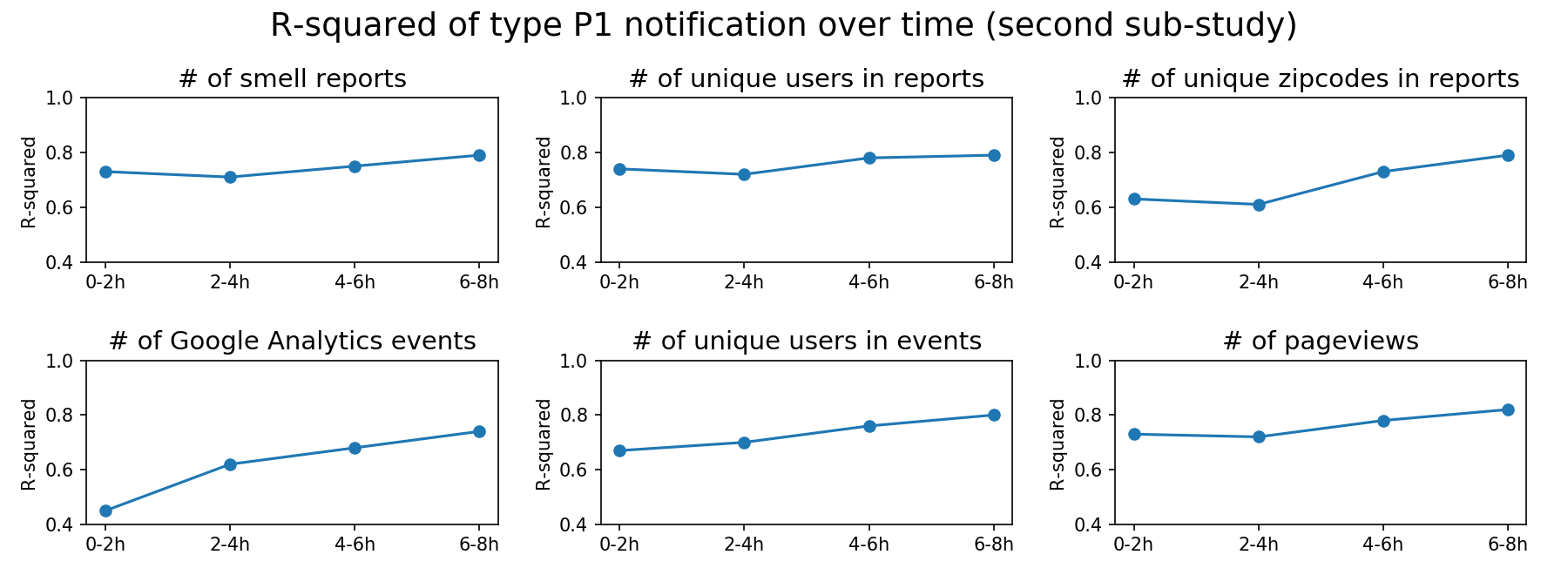}
	\caption{Analysis of the effect of type P1 smell predictive notifications overtime for the second sub-study.  The top, middle, and bottom parts demonstrate the trend of the magnitude (exponential of the treatment coefficient), significance (p-value of the treatment coefficient), and $R^2$ (effect size of the model) respectively. The title above each plot shows the corresponding metric. The x-axis means the bin of the time ranges after sending the notification. For instance, if a true positive happened at 8 am, \quo{2-4h} means the metric (e.g., number of Google Analytics events) measured from 10 am to 12 pm. This figure shows that after four hours, the magnitude of the effect drops to near 1 (i.e., no effect), and the treatment coefficient becomes statistically insignificant. The $R^2$ values are above 0.4 over the time ranges, which indicates reasonable goodness of fit.}
	\label{fig:time_tp_fn}
\end{figure}

One advantage of using GLM instead of transforming the data to fit a linear regression model is that we can conveniently interpret the meaning of the coefficients. For one unit increase in the $k^{th}$ predictor $X_k$, the response will be multiplied by $\exp(\hat{\beta_k})$ when using the natural log link function:
\begin{equation}
\tilde{Y} = \exp\Big(... + \beta_k (X_k+1) + ...\Big) = \exp(\beta_k) \cdot \exp \Big(... + \beta_k X_k + ...\Big) = \exp(\beta_k) \cdot Y
\end{equation}
where $\tilde{Y}$ is the increased response, and $Y$ is the original response. For instance, the treatment coefficient in the full model for the second sub-study is 1.16 when using the M4 metric (Table~\ref{tab:pn_analysis_tp_fn}), which means that we expect the number of Google Analytics events to be multiplied by $\exp(1.16)\approx3.19$ (exponential of coefficient) when the system sends a type P1 smell predictive notification, compared to the days with no notifications. Another example is that in the top graph of Figure~\ref{fig:time_tp_fn}, a higher exponential of coefficient means that the predictive push notification is more effective.

\begin{figure}[t]
	\centering
	\includegraphics[width=1\columnwidth]{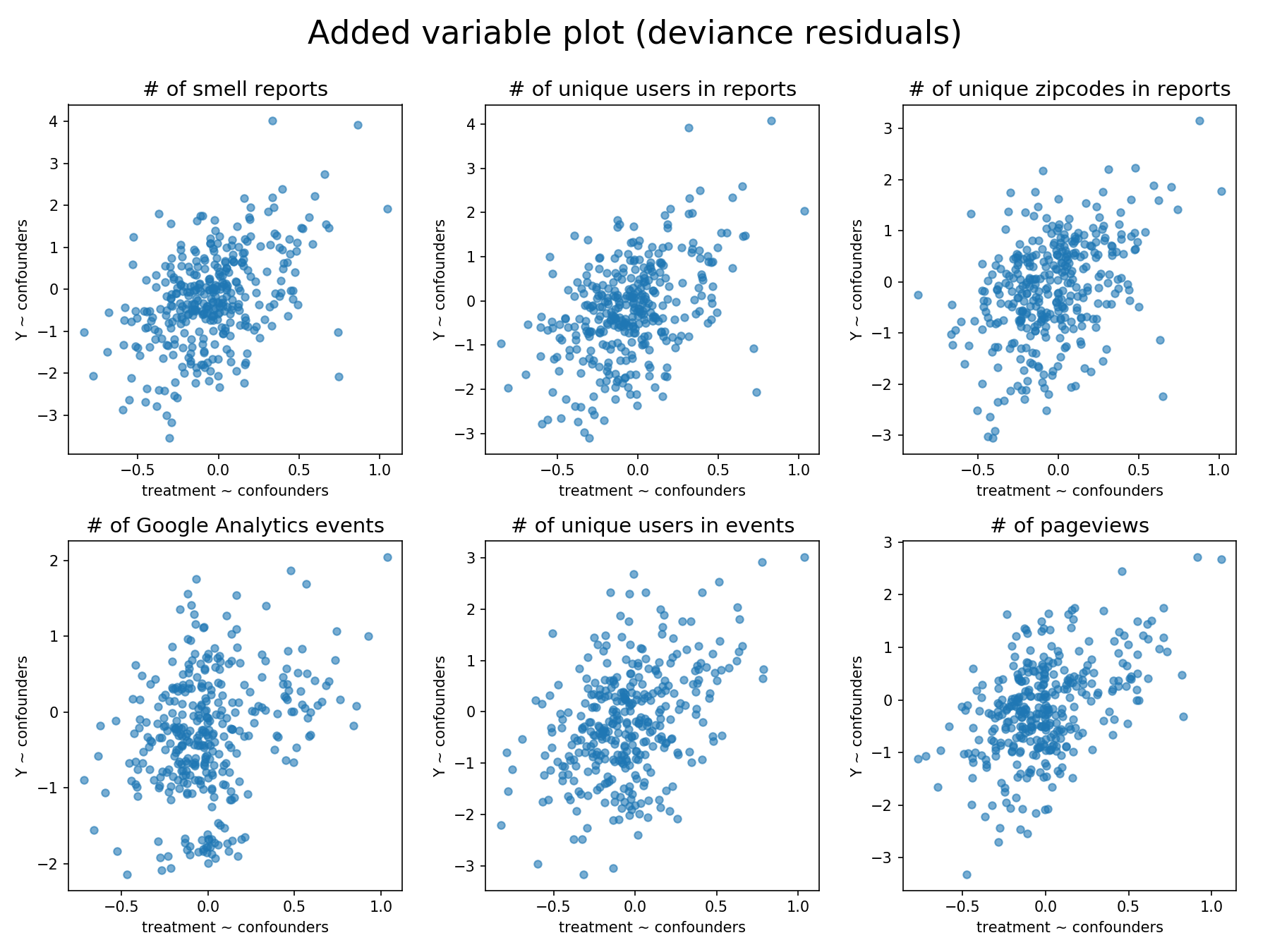}
	\caption{The added variable plots for the second sub-study. The x-axis shows deviance residuals of the treatment regressed on the confounders, which contains the new information contributed by the treatment. The y-axis shows deviance residuals of the response regressed on the confounders, which contains the unexplained information by the confounders. These plots indicate a positive linear trend between the treatment (type P1 smell predictive notification) and the measured metrics.}
	\label{fig:avp_tp_fn}
\end{figure}

\begin{table}[p]
	\caption{Analysis of the metrics (Table~\ref{tab:pn_type}) between two groups: within two hours after the \textbf{true positive (TP) and false negative (FN)}, which means predicted (and happened) smell events and non-predicted (but happened) smell events respectively. In the TP group, the system sent \textbf{type P1 push notifications} (\textit{Smell Event Alert: Local weather and pollution data indicates there may be a Pittsburgh smell event in the next few hours. Keep a nose out and report smells you notice!}). The \quo{FN} and \quo{TP} columns show \quo{median $\pm$ semi-interquartile range}. Symbol $\delta$ indicates Cliff's delta (effective size). Symbol $D$, $P$, and $R^2$ mean the deviance statistic, the Pearson statistic, and the pseudo R-squared respectively. Abbreviation \quo{Coef} and \quo{CI} mean the coefficient of treatment and the 95\% confidence interval respectively. The Z statistic, p-value, and CI come from the Wald test on the treatment coefficient. The \quo{Model} column shows the base model (with only the treatment) or the full model (with all the predictors and the elastic net regularization). The asterisk encoding of the \quo{Model} column indicates significance level p$<$.05 (*), p$<$.01 (**), or p$<$.001 (***). Symbol \quo{\checkmark} or \quo{...} means that the Wald test on the corresponding coefficient has p$<$.05 or p$\geq$.05 respectively. An empty cell or \quo{n/a} means that the variable is not selected. Predictors that have no data are not shown in the table columns.}
	\label{tab:pn_analysis_tp_fn}
	\begin{tabular}{lccccc}
\toprule
\multicolumn{6}{c}{\textbf{Mann-Whitney U test (FN and TP), N(FN)=281, N(TP)=69}} \\ 
Metric & FN & TP & U statistic & p-value & $\delta$ \\ 
\midrule
\rowcolor{lightgray} 
M1 & 3$\pm$2 & 16$\pm$10 & 1,587 & \textbf{<.001} & .84 \\
M2 & 3$\pm$2 & 15$\pm$10 & 1,598 & \textbf{<.001} & .84 \\
\rowcolor{lightgray} 
M3 & 2$\pm$2 & 10$\pm$3 & 1,460 & \textbf{<.001} & .85 \\
M4 & 12$\pm$12 & 123$\pm$68 & 1,174 & \textbf{<.001} & .88 \\
\rowcolor{lightgray} 
M5 & 2$\pm$2 & 13$\pm$7 & 854 & \textbf{<.001} & .91 \\
M6 & 7$\pm$4 & 39$\pm$21 & 666 & \textbf{<.001} & .93 \\
\bottomrule
\vspace{3pt}
\end{tabular}
\begin{tabular}{llccccccc}
\toprule
\multicolumn{9}{c}{\textbf{Regression analysis (FN and TP)}} \\ 
Metric & Model & $D$ & $P$ & $R^2$ & Coef & Z statistic & p-value & CI \\ 
\midrule
\rowcolor{lightgray} 
M1 & \textbf{base***} & 283 & 287 & .48 & 1.83 & 14.38 & \textbf{<.001} & [1.58, 2.07] \\
\rowcolor{lightgray} 
& \textbf{full***} & 364 & 325 & .73 & 1.17 & 10.22 & \textbf{<.001} & [0.94, 1.39] \\
M2 & \textbf{base***} & 279 & 283 & .48 & 1.83 & 14.36 & \textbf{<.001} & [1.58, 2.08] \\
& \textbf{full***} & 350 & 316 & .74 & 1.27 & 12.10 & \textbf{<.001} & [1.07, 1.48] \\
\rowcolor{lightgray} 
M3 & \textbf{base***} & 367 & 320 & .43 & 1.29 & 16.29 & \textbf{<.001} & [1.14, 1.45] \\
\rowcolor{lightgray} 
& \textbf{full***} & 309 & 258 & .63 & 0.82 & 7.71 & \textbf{<.001} & [0.61, 1.02] \\
M4 & \textbf{base***} & 295 & 246 & .32 & 1.97 & 10.33 & \textbf{<.001} & [1.60, 2.34] \\
& \textbf{full***} & 240 & 188 & .45 & 1.16 & 4.56 & \textbf{<.001} & [0.66, 1.66] \\
\rowcolor{lightgray} 
M5 & \textbf{base***} & 260 & 240 & .44 & 1.66 & 12.97 & \textbf{<.001} & [1.41, 1.92] \\
\rowcolor{lightgray} 
& \textbf{full***} & 360 & 347 & .67 & 1.18 & 12.12 & \textbf{<.001} & [0.99, 1.38] \\
M6 & \textbf{base***} & 230 & 216 & .53 & 1.74 & 14.16 & \textbf{<.001} & [1.50, 1.98] \\
& \textbf{full***} & 235 & 208 & .73 & 1.00 & 7.90 & \textbf{<.001} & [0.75, 1.25] \\
\bottomrule
\vspace{3pt}
\end{tabular}
\begin{tabular}{lccccccccccccccc}
\toprule
\multicolumn{16}{c}{\textbf{Selected confounders in the full model (FN and TP)}} \\ 
Metric & P2 & P3 & P4 & P5 & WD & O3 & PM & CO & SO2 & XH & YH & XM & YM & E & M\\ 
\midrule
\rowcolor{lightgray} 
M1 & ... &  &  &  & \checkmark & ... & ... & \checkmark & \checkmark & ... & \checkmark & \checkmark & \checkmark & \checkmark & \checkmark \\
M2 & \checkmark &  & ... & \checkmark & \checkmark & ... & ... & \checkmark & \checkmark & ... &  & ... & \checkmark & \checkmark & \checkmark \\
\rowcolor{lightgray} 
M3 &  &  &  &  & \checkmark & ... & ... & \checkmark & \checkmark & ... & \checkmark & ... & \checkmark & ... & \checkmark \\
M4 &  &  &  & \checkmark & \checkmark & \checkmark & ... & ... & ... &  & \checkmark & ... &  & ... & \checkmark \\
\rowcolor{lightgray} 
M5 & ... & \checkmark & ... &  & \checkmark & ... & ... & ... & \checkmark & \checkmark &  & \checkmark & \checkmark &  & \checkmark \\
M6 &  & \checkmark & ... &  & \checkmark & ... & ... & ... & \checkmark & ... & \checkmark & \checkmark & \checkmark & ... & \checkmark \\
\bottomrule
\end{tabular}
\end{table}

Besides fitting the full model as described in Equation~\ref{eq:full}, we also fit a base model that only involves the intercept and the treatment. Hence, by comparing the regression coefficient of the treatment in these two models, we can determine the confounding effect. An informal rule is that confounding exists if the changes in the treatment coefficient from the base to the full model is more than 10\%. Then, we can use an added variable plot (discussed in~\cite{Cook-1982}) to visualize the effect of the treatment on the response variable after confounding adjustment. The x-axis of the plot shows deviance residuals of the treatment regressed on the confounders, which contains the new information contributed by the treatment. The deviance residual $r_d$ is defined as
\begin{equation}
r_d = \mathrm{sign}(Y_i-\hat{\mu_i}) \cdot \sqrt{d_i}
\end{equation}
where $d_i$ is the unit contribution of observation $Y_i$ to the deviance ($D=\sum_{i=1}^n{d_i}$ in Equation~\ref{eq:deviance} in Appendix~\ref{ap:glm}). The y-axis is the deviance residual of the response regressed on the confounders, which contains the unexplained information by the confounders. Intuitively, the plot shows the relationship between the treatment and the response (i.e., measured metric) while holding all other confounders constant. For instance, the added variable plots (Figure~\ref{fig:avp_tp_fn}) for the second sub-study shows a positive linear trend between the response and the treatment after confounding adjustment.

To report the effect size of the regression analysis, we compute the deviance-based pseudo $R^2$ (in the range between 0 and 1), which generalizes the coefficient of determination for typical linear regression models~\cite{Cameron-1997}. It can be viewed as the proportion of uncertainty explained by the fitted model with the form
\begin{equation}
R^2 = 1 - D / D_{\mathrm{null}}
\end{equation}
where $D$ is the deviance of the fitted model (Equation~\ref{eq:deviance}), and $D_{\mathrm{null}}$ is the deviance of the null model (the one fitted with only the intercept). When fitting typical linear regression models with the Ordinary Least Squares algorithm, the pseudo $R^2$ is equivalent to the coefficient of determination that uses the residual sum of squares (RSS). Since GLM adopts maximum likelihood estimation through an iterative process, RSS cannot truly reflect the goodness of fit in our case. In general, a large $R^2$ value means that the model has a higher power to explain the uncertainty in the data, and the interpreted result can be more reliable. For instance, the $R^2$ for the full model with metric M6 in the second sub-study is $0.73$, and the treatment coefficient is statistically significant, which indicates strong evidence that the type P1 smell predictive notification is highly related to the changes in the Google Analytics pageviews. 

For the first sub-study, we report the results for each notification in Table~\ref{tab:pn_analysis_pn1}, \ref{tab:pn_analysis_pn2}, \ref{tab:pn_analysis_pn3}, \ref{tab:pn_analysis_pn4}, \ref{tab:pn_analysis_pn5}, and \ref{tab:pn_analysis_pn6} in Appendix~\ref{ap:pn_study_other_tables}. The result of the second sub-study for the type P1 push notification (smell prediction) is described in Table~\ref{tab:pn_analysis_tp_fn}. In sum, there is statistically significant evidence to indicate that type P1 and P3 push notifications are effective with reasonable large effect sizes, as shown in Table~\ref{tab:pn_type}. The magnitude (i.e., exponential of the coefficient) of the effect of type P1 notification drops dramatically to near 1 (i.e., no effect) after four hours, as shown in Figure~\ref{fig:time_tp_fn}.

\subsection{Survey Study}\label{sec:survey_study}

In this study, we show that the system can motivate active community members to contribute data and increase their self-efficacy, beliefs about how well an individual can achieve desired effects through actions \cite{Bandura-1977}. We recruited adult participants via snowball sampling \cite{Biernacki-1981}. We administered and delivered an anonymous online survey (Appendix~\ref{ap:survey_materials}) via email to community advocacy groups and asked them to distribute the survey to others. Paper surveys were also provided. All responses were kept confidential, and there was no compensation. We received 29 responses in total over one month from March 20th to April 20th, 2018. Four responses were excluded due to incomplete questions or no experiences in interacting with the system, which gave 25 valid survey responses. There were 8 males, 16 females, and 1 person with undisclosed gender information. All but one participant had a Bachelor's degree at minimum. The demographics of the sample population (Table \ref{tab:demographics}) were not typical for the region. The survey had three sections: (1) Self-Efficacy Changes, (2) Motivation Factors, (3) System Usage Information.

\begin{table}[t]
	\caption{Demographics of participants. Columns and rows represent ages and education levels.}
	\label{tab:demographics}
	\begin{tabular}{lccccccc}
		\toprule
		& 18-24\!\!\! & 25-34\!\!\! & 35-44\!\!\! & 45-54\!\!\! & 55-64\!\!\! & 65-74\!\!\! & Sum\!\!\!  \\
		\midrule
		Associate\!\!\!\! & 0 & 0 & 1 & 0 & 0 & 0 & 1 \\
		Bachelor\!\!\!\! & 2 & 2 & 2 & 0 & 1 & 1 & 8 \\
		Master\!\!\!\! & 0 & 2 & 2 & 0 & 0 & 4 & 8 \\
		Doctoral\!\!\!\! & 0 & 1 & 1 & 1 & 5 & 0 & 8  \\
		Sum\!\!\!\! & 2 & 5 & 6 & 1 & 6 & 5 & 25 \\
		\bottomrule
	\end{tabular}
\end{table}

For Self-Efficacy Changes, we measured changes to user confidence mitigating air quality problems. This section was framed as a retrospective pre-post self-assessment. The items were divided between pre-assessment, \quo{BEFORE you knew about or used Smell Pittsburgh,} and post-assessment, \quo{AFTER you knew about or used Smell Pittsburgh.} For both assessments, we used a scale developed by the Cornell Lab of Ornithology \cite{Porticella-2017-seea, DEVISE}. The scale was customized for air quality to suit our purpose. The scale consisted of eight Likert-type items (from 1 \quo{Strongly Disagree} to 5 \quo{Strongly Agree}).

\begin{table}[t]
	\caption{Frequency of system usage (sorted by percentage).}
	\label{tab:freq_of_usage}
	\begin{tabular}{lcc}
		\toprule
		& Count & Percentage \\
		\midrule
		Other (the open-response text field) & 9 & 36\%\\
		At least once per month & 7 & 28\%\\
		At least once per week & 4 & 16\%\\
		At least once per day & 3 & 12\%\\
		At least once per year & 2 & 8\%\\
		\bottomrule
	\end{tabular}
\end{table}

\begin{table}[t]
	\caption{The multiple-choice question for measuring participation level (sorted by percentage).}
	\label{tab:participation_level}
	\begin{tabular}{lcc}
		\toprule
		& Count \\
		\midrule
		I submitted smell reports. & 22 (88\%)\\
		I checked other people's smell reports on the map visualization. & 22 (88\%)\\
		I opened Smell Pittsburgh when I noticed unusual smell. & 22 (88\%)\\
		I discussed Smell Pittsburgh with other people. & 21 (84\%)\\
		I provided my contact information when submitting smell reports. & 14 (56\%)\\
		I paid attention to smell event alert notifications provided by Smell Pittsburgh. & 13 (52\%)\\
		I shared Smell Pittsburgh publicly online (e.g. email, social media, news blog). & 13 (52\%)\\
		I clicked on the playback button to view the animation of smell reports. & 9 (36\%)\\
		I took screenshots of Smell Pittsburgh. & 9 (36\%)\\
		I mentioned or presented Smell Pittsburgh to regulators. & 6 (24\%)\\
		I downloaded smell reports data from the Smell Pittsburgh website. & 4 (16\%)\\
		\bottomrule
	\end{tabular}
\end{table}

The Motivation Factors section was based on a scale developed by the Cornell Lab of Ornithology \cite{Porticella-2017-gene, DEVISE} with 14 Likert-type items (from 1 \quo{Strongly Disagree} to 5 \quo{Strongly Agree}). The scale was customized for air quality and measured both internal (7 items) and external motivations (7 items). Examples of internal motivations included enjoyment during participation and the desire to improve air quality. Examples of external motivations included the attempt to gain rewards and to avoid negative consequences if not taking actions. A text field with question \quo{Are there other reasons that you use Smell Pittsburgh?} was provided for open responses.

\begin{figure}[t]
	\centering
	\includegraphics[width=1\columnwidth]{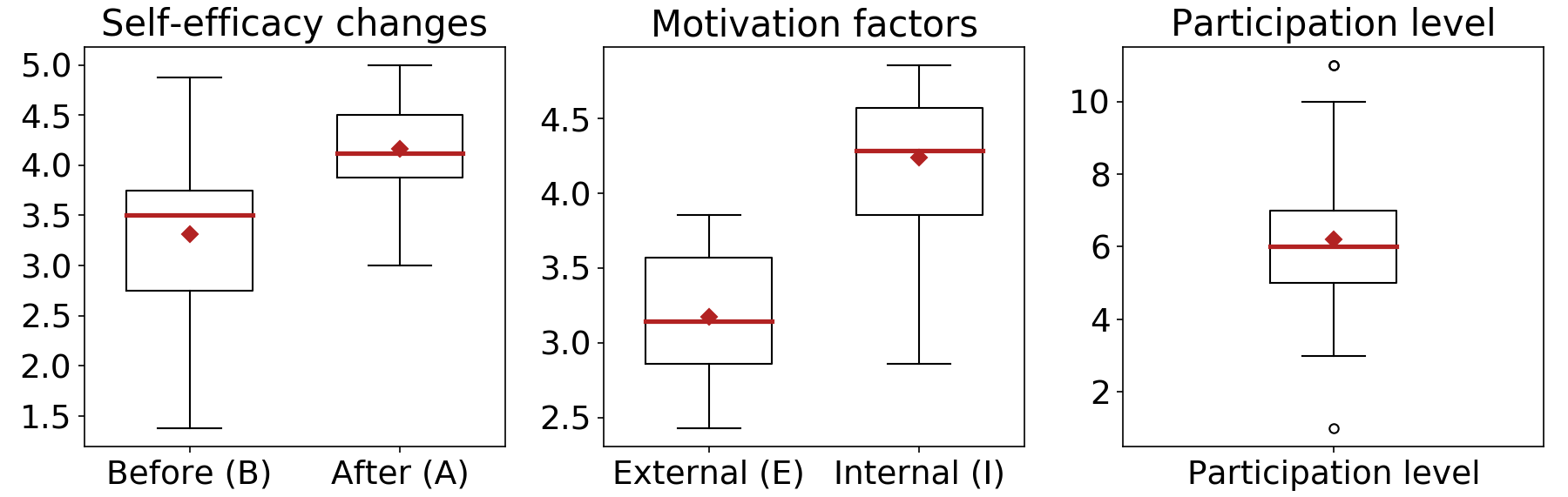}
	\caption{The distributions of self-efficacy changes, motivations, and participation level for our survey responses. The red lines in the middle of the box indicate the median. The red-filled diamonds represent the mean. The top and bottom edges of a box indicate 75\% ($Q3$) and 25\% ($Q1$) quantiles respectively. The boxes show inter-quantile ranges $IQR=Q3-Q1$. The top and bottom whiskers show $Q3+1.5*IQR$ and $Q1+1.5*IQR$ respectively. Black hollow circles show outliers.}
	\label{fig:survey}
\end{figure}

In the System Usage Information section, we collected individual experiences with Smell Pittsburgh. We documented participation level through a multiple-choice and multiple-response question, \quo{How did you use Smell Pittsburgh?} as shown in Figure \ref{fig:survey} (right). This question allowed participants to select from a list of 11 activities, which include submitting reports, interacting with the system, sharing experiences, and disseminating data (Table \ref{tab:participation_level}). We identified the frequency of system usage through a multiple-choice question, \quo{How often do you use Smell Pittsburgh?} as shown in Table \ref{tab:freq_of_usage}. Text fields were provided for both of the above two questions.

At the end of the survey, we asked an open-response question \quo{Do you have any other comments, questions, or concerns?} Our analysis is presented below along with each related question and selected quotes. Bold emphases in the quotes were added by researchers to highlight key user sentiments.

For Self-Efficacy, we averaged the scale items to produce total self-efficacy pre score (Mdn=3.50) and post score (Mdn=4.13) for each participant (Figure \ref{fig:survey}, left). A two-tailed Wilcoxon Signed-Ranks test (a nonparametric version of a paired t-test) indicated a statistically significant difference (T=13.5, Z=-3.79, p\textless.001, Spearman correlation=0.49). This finding indicated that there were increases in self-efficacy during participation.

For Motivation Factors, we averaged the internal (Mdn=4.29) and external (Mdn=3.14) motivation scores for each participant (Figure \ref{fig:survey}, center). A two-tailed Wilcoxon Signed-Ranks test indicated a statistically significant difference (T=0, Z=-4.29, p\textless.001, Spearman correlation=0.46). This result suggested that internal factors were primary motivations for our participants rather than external factors. Open-ended answers showed that nine participants (36\%) mentioned that the system enabled them to contribute data-driven evidence efficiently and intuitively.
\begin{quote}
	\quo{I used to try to use the phone to call in complaints, but that was highly unsatisfactory. I never knew if my complaints were even registered. With Smell Pittsburgh, \textbf{I feel that I'm contributing to taking data}, as well as to \textbf{complaining when it's awful}. [...]}
\end{quote}
\begin{quote}
	\quo{It's seems to be the most \textbf{effective way to report wood burning} that can fill my neighborhood with the smoke and emissions from wood burning.}
\end{quote}
\begin{quote}
	\quo{The Smell app \textbf{quantifies observations in real time.} Researchers can use this qualitative information along quantitative data in real time. Added benefit is to have [the health department] receive this information in real time \textbf{without having to make a phone call or send separate email}. I have confidence that the recording of Smell app data is \textbf{quantified more accurately} than [the health department]'s.}
\end{quote}
\begin{quote}
	\quo{It is an \textbf{evidence based way} for a citizen to register what is going on with the air where I live and work.}
\end{quote}
\begin{quote}
	\quo{I believe in science and data and think \textbf{this can help build a case}. [...]}
\end{quote}
Also, four participants (16\%) indicated the benefit to validate personal experiences based on the data provided by others.
\begin{quote}
	\quo{I used to (and sometimes still do) call reports in to [the health department]. I love how the map displays after I post a smell report. \textbf{Wow! I'm not alone!}}
\end{quote}
\begin{quote}
	\quo{It \textbf{validates my pollution experiences} because others are reporting similar experiences.}
\end{quote}
\begin{quote}
	\quo{I like using it for a similar reason that I like checking the weather. It helps me understand my environment and \textbf{confirms my sense of what I'm seeing}.}
\end{quote}
We also found that altruism, the concern about the welfare of others, was another motivation. Six participants (24\%) mentioned the desire to address climate changes, activate regulators, raise awareness of others, expand air quality knowledge, influence policy-making, and build a sense of community.
\begin{quote}
	\quo{Because \textbf{climate change} is one of our largest challenges, [...] Also, the ACHD isn't as active as they should be, and \textbf{needs a nudge}.}
\end{quote}
\begin{quote}
	\quo{I use [Smell Pittsburgh] to \textbf{demonstrate to others how they can raise their own awareness}. I've also pointed out to others that many who have grown up in this area of Western PA have grown up with so much pollution, \textbf{to them air pollution has become normalized} and many do not even smell the pollution any more. This is extremely dangerous and disturbing.}
\end{quote}
\begin{quote}
	\quo{I want to help \textbf{expand the knowledge and education of air quality} in Pittsburgh and believe the visuals Smell Pittsburgh provides is the best way to do that.}
\end{quote}
\begin{quote}
	\quo{I believe in the power of crowd-sourced data to \textbf{influence policy decisions}. I also believe that the air quality activism community will find more willing participants if there is a \textbf{very easy way for non-activists to help support clean air}, and the app provides that mechanism. It is basically a very \textbf{easy onramp for potential new activists}. The app also acts as a way for non-activists to see that they are not alone in their concerns about stinky air, which I believe was a major problem for \textbf{building momentum in the air quality community} prior to the app's existence.}
\end{quote}
For System Usage Information, we reported the counts for system usage frequency questions (Table \ref{tab:freq_of_usage}). The result showed that our users had a wide variety of system usage frequency. Open-responses indicated that instead of using the system regularly, eight participants (32\%) only submitted reports whenever they experienced poor odors. To quantify participation levels, we counted the number of selected choices for each participant, as shown in Figure \ref{fig:survey} (right). We found that our participation levels were normally distributed. In the open-response text field for this question, two participants (8\%) mentioned using personal resources to help promote the system.
\begin{quote}
	\quo{I \textbf{ran a Google Adwords campaign to get people to install Smell Pittsburgh}. It turns out that about \$6 of ad spending will induce someone to install the app.}
\end{quote}
\begin{quote}
	\quo{I take and share so many screenshots! Those are awesome. [...] I also \textbf{made two large posters of the app screen}-- one on a very bad day, and one on a very good day. I \textbf{bring them around to public meetings} and try to get county officials to look at them.}
\end{quote}
In the open-ended question to freely provide comments and concerns, two participants (8\%) were frustrated about the lack of responses from regulators and unclear values of using the data to take action.
\begin{quote}
	\quo{After using this app for over a year, and making many dozens of reports, I \textbf{haven't once heard from the [health department]}. That is disappointing, and makes me wonder, why bother? [...] Collecting this data is clever, but towards what end? I sometimes \textbf{don't see the point in continuing to report}.}
\end{quote}
\begin{quote}
	\quo{\textbf{It wasn't clear when using the app that my submission was counted} [...]. I want to be able to see directly that my smell reports are going somewhere and being used for something. [...]}
\end{quote}
Also, five (20\%) participants suggested augmenting the current system with new features and offering this mobile computing tool to more cities. Such features involved reporting smell retrospectively and viewing personal submission records.
\begin{quote}
	\quo{I \textbf{get around mostly by bike}, so it is difficult to report smells the same moment I smell them. I wish I could report smells in a different location than where I am so that I \textbf{could report the smell once I reach my destination}.}
\end{quote}
\begin{quote}
	\quo{It would be nice to be able to \textbf{add a retroactive report}. We often get the strong sulfur smells in Forest Hills in the middle of the night [...] but I strongly \textbf{prefer to not have to log in to my phone at 3 am to log the report} as it makes it harder to get back to sleep.}
\end{quote}
\begin{quote}
	\quo{This app should let me \textbf{see/download all of my data}: how many times I reported smells, what my symptoms and comments were and how many times the [health department] didn't respond [...]}
\end{quote}
	
	\section{Discussion}

We have shown that the smell data gathered through the system is practical in identifying local air pollution patterns. We released the system in September 2016. During 2017, the system collected 8,720 smell reports, which is 10-fold more than the 796 complaints collected by the health department regulators in 2016. All smell reports in our system had location data, while the location information was missing from 45\% of the regulator-collected complaints. Although there is a significant increase in data quantity (as described in the system usage study), researchers may criticize the reliability of the citizen-contributed data, since lay experiences may be subjective, inconsistent, and prone to noise. Despite these doubts, in the smell dataset study, we have applied machine learning to demonstrate the predictive and explainable power of these data. It is viable to forecast upcoming smell events based on previous observations. We also extracted connections between predictors and responses that reveal a dominant local air pollution pattern, which is a joint effect of hydrogen sulfide and wind information. This pattern could serve as hypotheses for future epidemiological studies. Since users tended not to report when there is no pollution odor, we recommend aggregating smell ratings by time to produce samples with no smell events. According to the experiments of different models, we suggest using the classification approach by thresholding smell ratings instead of the regression approach. In reality, the effect of 10 and 100 smell reports with high ratings may be the same for a local geographical region. It is highly likely that the regression function tried to fit the noise after a certain threshold that indicates the presence of a smell event.

Additionally, we have identified that certain types of push notifications are effective in user engagement (Table~\ref{tab:pn_type}). The push notifications study over 32 months of data indicates statistically significant evidence by using the regression analysis that takes the confounding variables into account. The result shows that the predictive notification (type P1) of odor problems can engage users to contribute more data (metric M1, M4, and M6) with a large effect size and notable increases in the median value (Table~\ref{tab:pn_analysis_tp_fn} and \ref{tab:pn_analysis_pn1}). The effect not only comes from more users (metric M2 and M5) but also a wider geographical region (metric M3), which indicates broader engagement. Moreover, the analysis over time demonstrates that predictive notifications remain effective for approximately four hours (Figure~\ref{fig:time_tp_fn}). Based on these findings, we recommend sending such types of predictive notifications to inform users about the potential occurrence of future events, which is analogous to the weather prediction. However, generating such notifications requires training a predictive model with statistical methods, which is not always viable at the early stage of system deployment due to lack of data. An alternative approach is to distribute the crowdsourced notifications (type P2 and P3) immediately when some users submit data or after collecting sufficient evidence. Despite smaller effect sizes for the type 6 push notification, we believe that such type of summary information (the total number of reports) can offer a sense of community around air quality concerns, which may be beneficial in the long term.

To analyze the effect of push notifications, we recommend using the generalized linear model (GLM) post hoc if conducting a randomized experiment is not feasible. When fitting the GLM, applying regularization techniques to select confounders can reduce multicollinearity. GLM allows researchers to make fewer assumptions about the data when compared to multiple linear regression. One may think of another approach, in which the dependent variable $Y$ is first transformed and then linearly regressed on the predictors $X$. However, this alternative estimates the conditional mean of the transformed $Y$ instead of the original $Y$, which means that explaining the result can be complicated because $Y$ and $X$ are no longer on the same scale. Moreover, in our case, the variance of measured metrics is unequal across the range of predictors' values (i.e., heteroscedasticity). It is difficult to find a transformation that can simultaneously improve linearity and stabilize the variance. By contrast, GLM is more flexible by modeling linearity and heteroscedasticity separately with the link and variance function, which can have a better potential to identify statistical significance.

We have also shown that the transparency of smell data empowered communities to advocate for better air quality. The findings in the survey study suggested that the system lowered the barrier for communities to contribute and communicate data-driven evidence. Although the small sample size limited the survey study, the result showed increases in self-efficacy after using the system. Several participants were even willing to use their resources to encourage others to install the system and engage in odor reporting. Moreover, in July 2018, activists attended the Board of Health meeting with the ACHD (Allegheny County Health Department) and presented a printed 230-foot-long scroll of more than 11,000 complaints submitted through the system. These reports allowed community members to ground their personal experiences with concrete and convincing data. The presented smell scroll demonstrated strong evidence about the impact of air pollution on the living quality of citizens, which forced regulators to respond to the air quality problem publicly. The deputy director of environmental health mentioned that ACHD would enact rigorous rules for coke plants: \highlightI{\quo{Every aspect of the activity and operation of these coke plants will have a more stringent standard applied \cite{smell-scroll-news-1-2018,smell-scroll-news-2-2018}.}} In this case, Smell Pittsburgh rebalanced the power relationships between communities and regulators.

\subsection{Limitation for the Survey Study}

We have described the design, deployment, and evaluation of a mobile smell reporting system for Pittsburgh communities to collect and visualize odor events. However, our survey study only targeted community activists, which led to a relatively small sample size. The commitment of these users may be driven by internal motivations, such as altruism, instead of the system. Additionally, our community members might be unique in their characteristics, such as the awareness of air pollution, the tenacity of advocacy, and the power relationships with other stakeholders. Involving citizens to address air pollution collaboratively is a wicked problem by its nature, so there is no guarantee that our success and effectiveness can be replicated in a different context. It is possible that interactive systems like Smell Pittsburgh can only be practical for communities with specific characteristics, such as high awareness. Future research is needed to study the impact of deploying this system in other cities that have similar or distinct community characteristics compared to Pittsburgh. It can also be beneficial to explore ways to connect citizens and regulators, such as visualizing smell reports by voting districts, providing background information with demographics and industry data, and sending push notifications regarding health agency public meetings.

\subsection{Limitation for the Smell Dataset Study}

From a machine-learning standpoint, community-powered projects such as Smell Pittsburgh often face two challenges that compromise model performances: data sparsity and label unreliability. Recent research has shown that deep neural networks can predict events effectively when equipped with a significant amount of training data~\cite{LeCun-2015}. However, the number of collected smell reports are far away from such level due to the limited size of the community and active users. The participated 3,917 users out of the 300,000 residents (1.3\%) is not sufficient to cover the entire Pittsburgh area. Additionally, in our case, air pollution incidents can only be captured at the moment because our communities lack resources to deploy reliable air quality monitoring sensors. It is impractical to annotate these incidents off-line such as in Galaxy Zoo~\cite{Raddick-2013}. While adopting transfer learning could take advantage of existing large-scale datasets from different domains to boost our performance~\cite{Pan-2010}, data sparsity is a nearly inevitable issue that must be taken into consideration for many community-powered projects.

Another issue is the label unreliability. There is no real \quo{ground truth} air pollution data in our case. The smell events defined in this research were based on the consensus from a group of users. As a consequence, the quality of the labels for the prediction and interpretation task could be influenced by confirmation bias, where people tend to search for information that confirms their prior beliefs. Such type of systematic error may be difficult to avoid, especially for community-powered projects. Future work to address these two challenges involves adding more predictors ({\em e.g.}, weather forecasts, air quality index) and using generalizable data interpretation techniques that can explain any predictive model to identify more patterns~\cite{Ribeiro-2016}.

\subsection{Limitation for the Push Notifications Study}

The findings in the associations between push notifications and measured metrics (e.g., number of Google Analytics events) do not imply causality. The push notifications analysis is an observational study, and we cannot conclude that certain types of notifications may cause differences in the measured metrics. Theoretically, we can precisely determine the effect of push notifications by conducting randomized experiments, such as A/B testing. However, under the Community Citizen Science (CCS) framework, conducting experiments that randomly assign users to the control or treatment group is extremely challenging and can have adverse impact. CCS research requires maintaining a sustainable relationship among stakeholders. In our case, controlling the information about air quality for different groups of people is strictly unethical from the community's point of view, which is against the goal of community empowerment and can break the mutual trust between communities and scientists. 

Moreover, the push notifications study may suffer from self-selection and self-assessment biases, even though the notifications have the same text descriptions without indicating specific odor types (e.g., rotten egg). For the self-selection bias, people who installed Smell Pittsburgh tend to have a high awareness of air pollution, and thus the samples do not represent the entire population. For the self-assessment bias, evaluations of smell ratings and odor severity, unlike sensors, can be subjective and vary by people. Thus, the trained machine learning model for predicting smell events is fundamentally biased. In the second sub-study, when comparing the true positives and false negatives in predictive push notifications, it is possible that the model only makes mistakes for certain types of odors since the training data come from self-reports. We mitigated the self-assessment biases by including possible confounding factors in the regression model (Table~\ref{tab:pn_predictors}).

Furthermore, the push notification analysis cannot reveal the real effect of type P4 and P5 notifications. These two types of notifications are designed to inform air quality changes, and one can imagine a situation that users look at the notification to gain information without opening the mobile application. Under this situation, the system is unable to measure user engagement quantitatively. Whether type P4 and P5 notifications are helpful remains an open research question. A future direction involves using a survey to gain feedback and understand usage patterns about push notifications. Another possible direction is to investigate if sending more push notifications is related to higher user engagement, and we leave this to future work.

\section{Conclusion}

This paper explores the design and impact of a mobile smell reporting system, Smell Pittsburgh, to empower communities in advocating for better air quality. The system enables citizens to submit and visualize odor experiences without the assistance from professionals. The visualization presents the context of air quality concerns from multiple perspectives as evidence. In our evaluation, we studied the distribution of smell reports and interaction events among different types of users. We also constructed a smell event dataset to study the value of these citizen-contributed data. By adopting machine learning, we developed a model to predict smell events and send push notifications accordingly. We also trained an explainable model to reveal connections between air quality sensor readings and smell events. The push notification study demonstrated that certain types of push notifications have statistically significant effects in increasing user engagement. Using a survey, we studied motivation factors for submitting smell reports and measured user attitude changes after using the system. Finally, we discussed limitations and future directions: deploying the system in multiple cities and using advanced techniques for pattern recognition. We envision that this research can inspire engineers, designers, and researchers to develop systems that support community advocacy and empowerment.
	
	\begin{acks}
		We thank the Heinz Endowments, the CREATE Lab (Jessica Pachuta, Ana Tsuhlares), Allegheny County Clean Air Now (ACCAN), PennEnvironment, Group Against Smog and Pollution (GASP), Sierra Club, Reducing Outdoor Contamination in Indoor Spaces (ROCIS), Blue Lens, PennFuture, Clean Water Action, Clean Air Council, the Global Communication Center of Carnegie Mellon University (Ryan Roderick), the Allegheny County Health Department, and all other participants for their collaboration.
	\end{acks}
	
	\bibliographystyle{ACM-Reference-Format}
	\bibliography{sample-base}
	
	\appendix
	
	\section{Appendix}

The appendix contains an overview and implementation details of the Generalized Linear Model for the push notification study, tables of the results for the push notifications study, and the materials for the survey study.

\subsection{Overview and Implementation Details of the Generalized Linear Model}\label{ap:glm}

Recall that in the push notifications study in section~\ref{sec:pn_study}, we aim to estimate the conditional mean $\epv(Y|X)$ of the Generalized Linear Model (GLM):
\begin{equation}
g\big(\epv(Y|X)\big) = \beta^T X = \beta_0 + \beta_1 X_1 + \beta_2 X_2 + ... + \beta_{m-1} X_{m-1}
\end{equation}
where $g$ is a link function, $\epv$ is the expected value, $\beta \in \real^{m \times 1}$ is the coefficient vector, $Y \in \real^{1 \times n}$ is the response vector, $X \in \real^{m \times n}$ is the predictor matrix, $m$ is the number of parameters (both predictors and intercept), and $n$ is the number of observations.

GLM treats the response $Y$ as a random variable that follows an exponential family distribution with probability density function $f$ in the form
\begin{equation}\label{eq:pdf}
f(y| \theta, \phi) = \exp \Bigg( \frac{y\theta - b(\theta)}{a(\phi)}  + c(y, \phi) \Bigg)
\end{equation}
where $y$ is an observation $Y_i$ of the random variable $Y$, $\exp$ is the natural exponential function, $\theta$ is the canonical parameter, $\phi$ is the dispersion parameter, and notation $a$, $b$, and $c$ are some functions. Notice that $\theta$, precisely $\theta(\mu)$, is a function of the conditional mean $\mu$. One convenient concept in GLM is that the conditional variance $V(Y|X)$ of the response variable can be expressed by the multiplication of the dispersion and a variance function $V(\mu)$ of the conditional mean $\mu$.
\begin{equation}
\begin{aligned}
\epv(Y|X) &= b'(\theta) = \mu = g^{-1}(\beta^T X) \\
V(Y|X) &= a(\phi) b''(\theta) = a(\phi)  V(\mu)
\end{aligned}
\end{equation}
In the basic linear regression, the link function $g(u) = u$ is an identify transformation, and $Y|X$ follows a normal distribution $N(\mu, \sigma^2)$ with mean $\mu$ and some variance $\sigma^2$. Hence, we have
\begin{equation}
\begin{aligned}
& \theta=\mu \mathcomma \phi=\sigma \mathcomma a(\phi)=\sigma^2\\
& \mu = \beta^T X \mathcomma V(\mu) = 1
\end{aligned}
\end{equation}
For normal distribution, the dispersion $a(\phi)$ is $\sigma^2$, and thus the variance function $V(\mu)$ is a constant. In our case, the response variable is not normally distributed. The metrics are strictly positive and have long tails on the right side, which means that the data is overdispersed with its variance larger than the mean. Moreover, a great number of measured metrics are near zero. This highly skewed pattern corresponds to the findings in the system usage study, in which only a few individuals contributed almost half of our data. Thus, we use a natural log link function $g(u) = \ln(u)$ and assume that $Y|X$ follows a negative binomial distribution $NB(\mu, \alpha)$ with mean $\mu$ and shape parameter $\alpha$.
\begin{equation}
\begin{aligned}
& \theta=\ln\Big(\frac{\alpha\mu}{1+\alpha\mu}\Big) \mathcomma \phi=1 \mathcomma a(\phi)=1\\
& \mu=\exp({\beta^T X}) \mathcomma V(\mu)=\mu+\alpha \mu^2
\end{aligned}
\end{equation}
where $\ln$ is the natural log function. In contrast to the normal distribution, the dispersion $a(\phi)$ is 1 for the negative binomial distribution. Using such distribution allows us to model the overdispersed response variable with the shape parameter $\alpha$ in the variance function. A reasonable $\alpha$ that fits our data can be chosen with model selection techniques (discussed later). When $\alpha=0$, the negative binomial distribution is reduced to a Poisson distribution.

GLM can be fitted with maximum likelihood estimation, which is equivalent to minimizing the negative log-likelihood. To reduce the number of predictors and simplify the model, we apply the elastic net regularization~\cite{Zou-2005}. Elastic net combines lasso ($L_1$ norm) and ridge ($L_2$ norm) penalties for selecting variables during optimization. Any predictor that has coefficient smaller than 0.00001 is removed, and the model is then re-fitted by using the remaining predictors. Notice that we do not penalize the intercept term $\beta_0$. Such combination grants the sparsity from lasso and stabilizes the coefficients with ridge, as shown in the following optimization goal:
\begin{equation}\label{eq:opt}
\hat{\beta} = \argmin_{\beta} \frac{-\ell(\theta)}{n} + \lambda \Big(w\norm{\beta}_1 + \frac{1-w}{2} \norm{\beta}^2_2\Big)
\end{equation}
where $n$ is the total number of observations (e.g., $n=2N$ for the analysis in Table~\ref{tab:pn_analysis_pn1}), $\lambda$ is the penalty, $w$ is the weight of lasso and ridge, and $\ell$ is the log-likelihood function of the probability density $f$ (Equation~\ref{eq:pdf}) with the following form:
\begin{equation}
\ell(\theta) = \ell(\theta, \phi| y) = \sum_{i=1}^{n} \Bigg( \frac{Y_i\theta - b(\theta)}{a(\phi)}  + c(Y_i, \phi) \Bigg)
\end{equation}
For simplicity, we choose $w=0.5$ throughout the study. To select the penalty $\lambda$ and shape parameter $\alpha$, we perform a grid search ($\lambda \in \{0.01, 0.1, 1\},\; \alpha \in \{0, 0.1, 0.2, 0.4, 0.8, 1, 2, 4, 8\}$) and choose the model with the lowest Akaike Information Criterion score~\cite{Akaike-1974}
\begin{equation}
\mathrm{AIC} = 2m - 2\ell(\hat{\theta})
\end{equation}
where $\hat{\theta}$ is the maximum likelihood estimate of the canonical parameter in the fitted model, $\ell(\hat{\theta})$ is the maximized log-likelihood, and $m$ is the number of parameters (both predictors and intercept) in the model. Notice that $\hat{\theta} = \hat{\theta}(\hat{\mu})$ is a function of the estimated conditional mean $\hat{\mu} = g^{-1}(\hat{\beta}^T X)$, which is assumed to be a transformation of the linear combination of the estimated coefficients $\hat{\beta}$. AIC penalizes complex models and rewards better goodness of fit (the negative log-likelihood to minimize in Equation~\ref{eq:opt}). Asymptotically when $n$ reaches infinity, AIC is equivalent to the leave-one-out cross-validation statistic~\cite{Stone-1977}.

To evaluate the fitted model, we first check if multicollinearity exists in our data. Multicollinearity is a phenomenon that some predictor variables are highly related to a linear combination of the others. Although multicollinearity does not affect how well the model fits the data, it may introduce difficulties while interpreting the result. Such difficulties include unstable coefficients during variable selection, large standard error and confidence interval for the coefficients, and different conclusions when performing hypothesis testing to determine if a coefficient is non-zero. To detect multicollinearity, we compute the Variance Inflation Factors (VIF) of the predictors using
\begin{equation}
\mathrm{VIF}_k = 1 / (1-R_k^2)
\end{equation}
where $\mathrm{VIF}_k$ is the variance inflation factor of the $k^{th}$ predictor, and $R_k^2$ is the coefficient of determination of predictor $X_k$ linearly regressed on all other predictors. In other words, $\mathrm{VIF}_k$ reflects the r-squared value of the regression model
\begin{equation}
X_k = a_0 + \sum_{j=1, j\neq k}^{m-1} a_j X_j
\end{equation}
In general, $\mathrm{VIF}>10$ indicates severe multicollinearity. For both the first and second sub-study, most of the predictors selected by the elastic net regularization have VIFs smaller than 4. An exception is that the VIFs for the cyclical transformations for time (XH and YH) are higher and can reach up to 9, which indicates moderate high multicollinearity for some fitted models (the fourth bin from 6 to 8 hours after sending type P1 notification in the second sub-study).

To select models that fit the data well, we conduct the goodness of fit test by comparing the regressed GLM model $\hat{M}$ and the saturated model $M^*$. The conditional means of $\hat{M}$ and $M^*$ are the maximum likelihood estimate $\hat{\mu}$ and the observed value $Y$ respectively. In other words, the saturated model is the best possible model when the predictions are the same as the observations. The test statistic that we used, deviance $D$, is defined as
\begin{equation}\label{eq:deviance}
D = -2 \Big(\ell(\hat{\theta}) - \ell(\theta^*) \Big) = -2 \sum_{i=1}^n \frac{Y_i\hat{\theta}_i - b(\hat{\theta}_i) - Y_i\theta^*_i + b(\theta^*_i)}{a(\phi)} \sim \chi^2_{n-m}
\end{equation}
where $i$ is the $i^{th}$ observation, and $\theta^* = \theta^*(\mu^*) = \theta^*(Y)$ is the canonical parameter in the saturated model. Deviance is a measure of the distance between the probability distributions of the fitted model $\hat{M}$ and the saturated model $M^*$, which is proportional to the Kullback-Leibler divergence~\cite{Hastie-1987}. Intuitively, the distance should be close to zero asymptotically with no more than the normally distributed sampling error. When using deviance as the test statistic, the goodness of fit test is equivalent to a likelihood ratio test between $\hat{M}$ and $M^*$. According to Wilks' theorem~\cite{Wilks-1938}, the likelihood ratio test statistic ($D$ in our case) follows a chi-squared distribution asymptotically with degrees of freedom $n-m$ under the null hypothesis, where $n$ and $m$ are the degrees of freedom (DOF) of the saturated and fitted model respectively. DOF describes the effective number of parameters, a quantitative measure of the model complexity. In the saturated model, DOF is the number of observations, which means that each observation has a corresponding regression coefficient.

The null hypothesis of the goodness of fit test is that the regressed GLM model $\hat{M}$ fits the data well, which implies that the deviance is smaller than the $\chi^2_{n-m}$ critical value of the desired significance level. In our case, we choose 0.1 as the significance level during model selection. In other words, the p-values of the goodness of fit test for all our fitted models are greater than 0.1, which means no rejections to the null hypothesis.

Besides using deviance for the goodness of fit test, we also apply the Pearson statistic $P$
\begin{equation}
P = \frac{S(\hat{\theta})^2}{I(\hat{\theta})} = \frac{\Big(\nabla\ell(\hat{\theta})\Big)^2}{-\nabla^2\ell(\hat{\theta})} =  \sum_{i=1}^{n} \frac{(Y_i - \hat{\mu_i})^2}{V(\hat{\mu_i})} \sim \chi^2_{n-m}
\end{equation}
where $S$ is the score function (slope of the log-likelihood function), $I$ is the Fisher information (variance of the score function), $\nabla$ is the partial derivative of the $\theta$ parameter, and $\hat{\mu_i} = g^{-1}(\hat{\beta}^T X_i)$ is the estimated conditional mean of the $i^{th}$ observation $Y_i$. When using the Pearson statistic, the goodness of fit test is equivalent to a score test (also known as the Lagrange multiplier test~\cite{Breusch-1980,Silvey-1959}) between the fitted model and the saturated model~\cite{Smyth-2003}. The intuition is that the slope (first partial derivative) at the estimated $\hat{\theta}$ of the log-likelihood function $\ell$ should be asymptotically close to zero with no more than the normally distributed sampling error when $\ell$ is maximized. In other words, the score test statistic ($P$ in our case) follows a chi-squared distribution asymptotically with degrees of freedom $n-m$ under the null hypothesis, as shown in the prior research~\cite{Rao-1948}. During model selection, we also choose 0.1 as the significance level for this score test.

To identify the effect of the treatment and confounders, we apply a zero-coefficient test with the Wald statistic. The null hypothesis is that the tested coefficient $\beta_k$ for the $k^{th}$ predictor $X_k$ is zero, which means no effect. The one-parameter Wald statistic is defined as
\begin{equation}
Z = \frac{\hat{\beta_i}}{\sqrt{V(\hat{\beta_i})}} \sim N(0, 1)
\end{equation}
where $V(\hat{\beta_k})$ is the variance of $\hat{\beta_k}$ computed during maximum likelihood estimation. The Wald test statistic $Z$ follows a standard normal distribution asymptotically under the null hypothesis~\cite{Wald-1943}. Different from the goodness of fit test that compares the fitted model and the saturated model, this zero-coefficient test compares two nested models: the one with all coefficients and the one that excludes $\beta_k$. A low p-value (<.05)  indicates that the effect of the corresponding predictor $X_k$ (e.g., treatment) is statistically significant. In other words, changes in $X_k$ are associated with changes in the response variable $Y$, and there is evidence that we should keep $X_k$ in the model. For instance, the p-values of the treatment coefficient in the second sub-study are all statistically significant (Table~\ref{tab:pn_analysis_tp_fn}), which indicates that type P1 push notification is effective.

\subsection{Other Tables for the Results of the Push Notifications Study}\label{ap:pn_study_other_tables}

The following pages contain tables that show the result of the push notifications study in section~\ref{sec:pn_study}.

\begin{table}[p]
	\caption{Analysis of the metrics (Table~\ref{tab:pn_type}) between two groups: within two hours \textbf{before and after} sending \textbf{type P1 notifications} (\textit{Smell Event Alert: Local weather and pollution data indicates there may be a Pittsburgh smell event in the next few hours. Keep a nose out and report smells you notice!}). The \quo{Before} and \quo{After} columns show \quo{median $\pm$ semi-interquartile range}. Symbol $\rho$ is the Spearman's correlation coefficient (effect size). Other notations and abbreviations in this table are explained in Table~\ref{tab:pn_analysis_tp_fn}.}
	\label{tab:pn_analysis_pn1}
	\begin{tabular}{lccccc}
\toprule
\multicolumn{6}{c}{\textbf{Wilcoxon signed-rank test (notification type P1), N=71}} \\ 
Metric & Before & After & T statistic & p-value & $\rho$ \\ 
\midrule
\rowcolor{lightgray} 
M1 & 7$\pm$7 & 16$\pm$10 & 676 & \textbf{.002} & .44 \\
M2 & 6$\pm$7 & 15$\pm$10 & 660 & \textbf{.002} & .46 \\
\rowcolor{lightgray} 
M3 & 5$\pm$4 & 10$\pm$3 & 386 & \textbf{<.001} & .51 \\
M4 & 43$\pm$45 & 117$\pm$68 & 304 & \textbf{<.001} & .57 \\
\rowcolor{lightgray} 
M5 & 5$\pm$5 & 13$\pm$7 & 290 & \textbf{<.001} & .60 \\
M6 & 14$\pm$15 & 39$\pm$21 & 298 & \textbf{<.001} & .58 \\
\bottomrule
\vspace{3pt}
\end{tabular}
\begin{tabular}{llccccccc}
\toprule
\multicolumn{9}{c}{\textbf{Regression analysis (notification type P1)}} \\ 
Metric & Model & $D$ & $P$ & $R^2$ & Coef & Z statistic & p-value & CI \\ 
\midrule
\rowcolor{lightgray} 
M1 & \textbf{base*} & 145 & 160 & .03 & 0.35 & 2.03 & \textbf{.043} & [0.01, 0.69] \\
\rowcolor{lightgray} 
& \textbf{full***} & 139 & 121 & .57 & 0.60 & 5.06 & \textbf{<.001} & [0.36, 0.83] \\
M2 & \textbf{base*} & 144 & 160 & .03 & 0.35 & 2.04 & \textbf{.041} & [0.01, 0.69] \\
& \textbf{full***} & 132 & 115 & .59 & 0.59 & 5.03 & \textbf{<.001} & [0.36, 0.83] \\
\rowcolor{lightgray} 
M3 & \textbf{base**} & 131 & 121 & .07 & 0.38 & 3.13 & \textbf{.002} & [0.14, 0.62] \\
\rowcolor{lightgray} 
& \textbf{full***} & 128 & 113 & .58 & 0.44 & 5.36 & \textbf{<.001} & [0.28, 0.60] \\
M4 & \textbf{base***} & 154 & 124 & .10 & 0.72 & 4.29 & \textbf{<.001} & [0.39, 1.05] \\
& \textbf{full***} & 110 & 100 & .48 & 0.95 & 6.26 & \textbf{<.001} & [0.65, 1.25] \\
\rowcolor{lightgray} 
M5 & \textbf{base***} & 118 & 124 & .11 & 0.59 & 3.75 & \textbf{<.001} & [0.28, 0.90] \\
\rowcolor{lightgray} 
& \textbf{full***} & 131 & 128 & .64 & 0.74 & 7.83 & \textbf{<.001} & [0.55, 0.92] \\
M6 & \textbf{base***} & 131 & 144 & .10 & 0.60 & 3.91 & \textbf{<.001} & [0.30, 0.90] \\
& \textbf{full***} & 94 & 87 & .66 & 0.85 & 7.54 & \textbf{<.001} & [0.62, 1.06] \\
\bottomrule
\vspace{3pt}
\end{tabular}
\begin{tabular}{lcccccccccccccc}
\toprule
\multicolumn{15}{c}{\textbf{Selected confounders in the full model (notification type P1)}} \\ 
Metric & P2 & P4 & P5 & WD & O3 & PM & CO & SO2 & XH & YH & XM & YM & E & M\\ 
\midrule
\rowcolor{lightgray} 
M1 & ... & ... & ... & \checkmark & ... & \checkmark & ... & ... & \checkmark & \checkmark & ... &  & ... & \checkmark \\
M2 & ... &  &  & \checkmark & ... & \checkmark & ... & ... & \checkmark & \checkmark & ... & ... & ... & \checkmark \\
\rowcolor{lightgray} 
M3 &  & ... & ... & \checkmark &  & ... & \checkmark & ... & \checkmark & \checkmark & \checkmark &  & ... & \checkmark \\
M4 &  &  &  &  & ... & ... & ... & \checkmark &  & \checkmark &  &  & \checkmark & \checkmark \\
\rowcolor{lightgray} 
M5 &  &  &  &  &  & \checkmark & ... & \checkmark & ... & \checkmark & ... &  & ... & \checkmark \\
M6 &  &  &  & \checkmark & ... & \checkmark & ... & \checkmark &  & \checkmark & ... &  &  & \checkmark \\
\bottomrule
\end{tabular}
\end{table}

\begin{table}[p]
	\caption{Analysis of the metrics (Table~\ref{tab:pn_type}) between two groups: within two hours \textbf{before and after} sending \textbf{type P2 notifications} (\textit{Smell Event Alert: Many residents are reporting poor odors in Pittsburgh. Were you affected by this smell event? Be sure to submit a smell report!}). The \quo{Before} and \quo{After} columns show \quo{median $\pm$ semi-interquartile range}. Symbol $\rho$ is the Spearman's correlation coefficient (effect size). Other notations and abbreviations in this table are explained in Table~\ref{tab:pn_analysis_tp_fn}.}
	\label{tab:pn_analysis_pn2}
	\begin{tabular}{lccccc}
\toprule
\multicolumn{6}{c}{\textbf{Wilcoxon signed-rank test (notification type P2), N=79}} \\ 
Metric & Before & After & T statistic & p-value & $\rho$ \\ 
\midrule
\rowcolor{lightgray} 
M1 & 8$\pm$8 & 8$\pm$4 & 1,300 & .637 & .45 \\
M2 & 7$\pm$8 & 8$\pm$4 & 1,258 & .611 & .47 \\
\rowcolor{lightgray} 
M3 & 6$\pm$4 & 6$\pm$2 & 1,106 & .325 & .46 \\
M4 & 72$\pm$38 & 105$\pm$55 & 628 & \textbf{<.001} & .40 \\
\rowcolor{lightgray} 
M5 & 9$\pm$4 & 13$\pm$5 & 600 & \textbf{<.001} & .45 \\
M6 & 25$\pm$13 & 34$\pm$10 & 903 & \textbf{.004} & .45 \\
\bottomrule
\vspace{3pt}
\end{tabular}
\begin{tabular}{llccccccc}
\toprule
\multicolumn{9}{c}{\textbf{Regression analysis (notification type P2)}} \\ 
Metric & Model & $D$ & $P$ & $R^2$ & Coef & Z statistic & p-value & CI \\ 
\midrule
\rowcolor{lightgray} 
M1 & base & 135 & 152 & .00 & -0.11 & -0.65 & .516 & [-0.43, 0.22] \\
\rowcolor{lightgray} 
& full & 141 & 123 & .50 & -0.01 & -0.05 & .961 & [-0.23, 0.22] \\
M2 & base & 133 & 147 & .00 & -0.11 & -0.65 & .513 & [-0.43, 0.22] \\
& full & 135 & 118 & .52 & n/a & n/a & n/a & n/a \\
\rowcolor{lightgray} 
M3 & base & 141 & 129 & .00 & 0.07 & 0.63 & .531 & [-0.16, 0.30] \\
\rowcolor{lightgray} 
& full & 144 & 124 & .51 & 0.11 & 1.35 & .177 & [-0.05, 0.26] \\
M4 & \textbf{base**} & 140 & 141 & .05 & 0.40 & 2.82 & \textbf{.005} & [0.12, 0.68] \\
& \textbf{full***} & 159 & 130 & .44 & 0.47 & 4.62 & \textbf{<.001} & [0.27, 0.67] \\
\rowcolor{lightgray} 
M5 & \textbf{base**} & 149 & 168 & .04 & 0.29 & 2.65 & \textbf{.008} & [0.08, 0.51] \\
\rowcolor{lightgray} 
& \textbf{full***} & 126 & 117 & .51 & 0.35 & 4.02 & \textbf{<.001} & [0.18, 0.51] \\
M6 & base & 82 & 96 & .03 & 0.21 & 1.45 & .146 & [-0.07, 0.49] \\
& \textbf{full***} & 140 & 133 & .53 & 0.30 & 3.92 & \textbf{<.001} & [0.15, 0.45] \\
\bottomrule
\vspace{3pt}
\end{tabular}
\begin{tabular}{lcccccccccccccc}
\toprule
\multicolumn{15}{c}{\textbf{Selected confounders in the full model (notification type P2)}} \\ 
Metric & P1 & P4 & P5 & WD & O3 & PM & CO & SO2 & XH & YH & XM & YM & E & M\\ 
\midrule
\rowcolor{lightgray} 
M1 & ... & ... & ... &  & ... & ... & ... & \checkmark & \checkmark & \checkmark & ... & ... & ... & \checkmark \\
M2 & ... & ... & ... &  & ... & ... & ... & \checkmark & \checkmark & \checkmark & ... & ... & ... & \checkmark \\
\rowcolor{lightgray} 
M3 &  &  &  &  & ... & \checkmark & ... & ... & \checkmark & \checkmark & ... & \checkmark &  & \checkmark \\
M4 &  & ... &  & ... &  & ... & ... & ... & \checkmark & \checkmark & ... & ... &  & \checkmark \\
\rowcolor{lightgray} 
M5 & ... &  & ... & ... &  & ... & ... & ... & \checkmark & \checkmark & ... &  & ... & \checkmark \\
M6 & ... & ... & ... & ... & ... & ... & ... & ... & \checkmark & \checkmark & ... & ... & ... & \checkmark \\
\bottomrule
\end{tabular}
\end{table}

\begin{table}[p]
	\caption{Analysis of the metrics (Table~\ref{tab:pn_type}) between two groups: within two hours \textbf{before and after} sending \textbf{type P3 notifications} (\textit{How does your air smell? A smell report rated 5 was just submitted.}). The \quo{Before} and \quo{After} columns show \quo{median $\pm$ semi-interquartile range}. Symbol $\rho$ is the Spearman's correlation coefficient (effect size). Other notations and abbreviations in this table are explained in Table~\ref{tab:pn_analysis_tp_fn}.}
	\label{tab:pn_analysis_pn3}
	\begin{tabular}{lccccc}
\toprule
\multicolumn{6}{c}{\textbf{Wilcoxon signed-rank test (notification type P3), N=377}} \\ 
Metric & Before & After & T statistic & p-value & $\rho$ \\ 
\midrule
\rowcolor{lightgray} 
M1 & 1$\pm$1 & 4$\pm$3 & 1,768 & \textbf{<.001} & .56 \\
M2 & 1$\pm$1 & 4$\pm$2 & 1,309 & \textbf{<.001} & .56 \\
\rowcolor{lightgray} 
M3 & 1$\pm$1 & 4$\pm$2 & 1,106 & \textbf{<.001} & .56 \\
M4 & 4$\pm$7 & 19$\pm$20 & 7,848 & \textbf{<.001} & .36 \\
\rowcolor{lightgray} 
M5 & 1$\pm$1 & 4$\pm$2 & 2,680 & \textbf{<.001} & .44 \\
M6 & 3$\pm$2 & 15$\pm$8 & 722 & \textbf{<.001} & .57 \\
\bottomrule
\vspace{3pt}
\end{tabular}
\begin{tabular}{llccccccc}
\toprule
\multicolumn{9}{c}{\textbf{Regression analysis (notification type P3)}} \\ 
Metric & Model & $D$ & $P$ & $R^2$ & Coef & Z statistic & p-value & CI \\ 
\midrule
\rowcolor{lightgray} 
M1 & \textbf{base***} & 546 & 713 & .19 & 1.29 & 11.54 & \textbf{<.001} & [1.07, 1.50] \\
\rowcolor{lightgray} 
& \textbf{full***} & 730 & 681 & .73 & 1.28 & 21.56 & \textbf{<.001} & [1.17, 1.40] \\
M2 & \textbf{base***} & 527 & 660 & .20 & 1.30 & 11.58 & \textbf{<.001} & [1.08, 1.51] \\
& \textbf{full***} & 705 & 639 & .72 & 1.28 & 21.29 & \textbf{<.001} & [1.16, 1.40] \\
\rowcolor{lightgray} 
M3 & \textbf{base***} & 681 & 635 & .22 & 1.12 & 13.76 & \textbf{<.001} & [0.96, 1.28] \\
\rowcolor{lightgray} 
& \textbf{full***} & 671 & 585 & .65 & 1.14 & 20.24 & \textbf{<.001} & [1.03, 1.26] \\
M4 & \textbf{base***} & 521 & 532 & .09 & 1.09 & 7.45 & \textbf{<.001} & [0.81, 1.38] \\
& \textbf{full***} & 681 & 612 & .34 & 1.16 & 11.00 & \textbf{<.001} & [0.95, 1.37] \\
\rowcolor{lightgray} 
M5 & \textbf{base***} & 717 & 751 & .23 & 1.16 & 14.60 & \textbf{<.001} & [1.00, 1.32] \\
\rowcolor{lightgray} 
& \textbf{full***} & 725 & 648 & .60 & 1.18 & 20.02 & \textbf{<.001} & [1.07, 1.30] \\
M6 & \textbf{base***} & 648 & 770 & .35 & 1.32 & 18.88 & \textbf{<.001} & [1.19, 1.46] \\
& \textbf{full***} & 549 & 496 & .67 & 1.36 & 24.84 & \textbf{<.001} & [1.25, 1.47] \\
\bottomrule
\vspace{3pt}
\end{tabular}
\begin{tabular}{lccccccccccccc}
\toprule
\multicolumn{14}{c}{\textbf{Selected confounders in the full model (notification type P3)}} \\ 
Metric & P4 & P5 & WD & O3 & PM & CO & SO2 & XH & YH & XM & YM & E & M\\ 
\midrule
\rowcolor{lightgray} 
M1 &  & ... & \checkmark & ... & \checkmark & ... & ... & \checkmark & ... &  & \checkmark & \checkmark & \checkmark \\
M2 &  &  &  & ... & \checkmark & ... & ... & ... & ... &  & \checkmark & \checkmark & \checkmark \\
\rowcolor{lightgray} 
M3 &  &  &  & \checkmark & ... & \checkmark & \checkmark &  &  &  & ... & \checkmark & \checkmark \\
M4 &  &  &  & ... & \checkmark & \checkmark & ... &  &  &  &  &  & \checkmark \\
\rowcolor{lightgray} 
M5 &  &  &  & \checkmark & ... & ... & \checkmark &  &  &  &  & \checkmark & \checkmark \\
M6 &  &  & \checkmark & ... & \checkmark & ... & ... & ... & \checkmark &  & \checkmark & ... & \checkmark \\
\bottomrule
\end{tabular}
\end{table}

\begin{table}[p]
	\caption{Analysis of the metrics (Table~\ref{tab:pn_type}) between two groups: within two hours \textbf{before and after} sending \textbf{type P4 notifications} (\textit{Does it smell better? Pittsburgh AQI just improved.}). The \quo{Before} and \quo{After} columns show \quo{median $\pm$ semi-interquartile range}. Symbol $\rho$ is the Spearman's correlation coefficient (effect size). Other notations and abbreviations in this table are explained in Table~\ref{tab:pn_analysis_tp_fn}.}
	\label{tab:pn_analysis_pn4}
	\begin{tabular}{lccccc}
\toprule
\multicolumn{6}{c}{\textbf{Wilcoxon signed-rank test (notification type P4), N=753}} \\ 
Metric & Before & After & T statistic & p-value & $\rho$ \\ 
\midrule
\rowcolor{lightgray} 
M1 & 1$\pm$2 & 1$\pm$2 & 54,082 & \textbf{<.001} & .37 \\
M2 & 1$\pm$1 & 1$\pm$2 & 52,558 & \textbf{<.001} & .37 \\
\rowcolor{lightgray} 
M3 & 1$\pm$1 & 1$\pm$2 & 48,876 & \textbf{<.001} & .38 \\
M4 & 12$\pm$16 & 12$\pm$16 & 112,080 & .361 & .42 \\
\rowcolor{lightgray} 
M5 & 2$\pm$2 & 2$\pm$2 & 82,362 & .065 & .47 \\
M6 & 6$\pm$4 & 8$\pm$6 & 93,096 & \textbf{<.001} & .54 \\
\bottomrule
\vspace{3pt}
\end{tabular}
\begin{tabular}{llccccccc}
\toprule
\multicolumn{9}{c}{\textbf{Regression analysis (notification type P4)}} \\ 
Metric & Model & $D$ & $P$ & $R^2$ & Coef & Z statistic & p-value & CI \\ 
\midrule
\rowcolor{lightgray} 
M1 & base & 481 & 848 & .01 & 0.27 & 1.78 & .075 & [-0.03, 0.56] \\
\rowcolor{lightgray} 
& \textbf{full**} & 1,399 & 1,388 & .34 & 0.19 & 3.03 & \textbf{.002} & [0.07, 0.31] \\
M2 & base & 474 & 848 & .01 & 0.28 & 1.87 & .062 & [-0.01, 0.57] \\
& \textbf{full**} & 1,354 & 1,314 & .35 & 0.20 & 3.21 & \textbf{.001} & [0.08, 0.32] \\
\rowcolor{lightgray} 
M3 & \textbf{base***} & 1,501 & 1,444 & .01 & 0.24 & 3.69 & \textbf{<.001} & [0.11, 0.36] \\
\rowcolor{lightgray} 
& \textbf{full***} & 1,460 & 1,394 & .33 & 0.21 & 3.82 & \textbf{<.001} & [0.10, 0.31] \\
M4 & base & 1,176 & 1,017 & .00 & 0.01 & 0.12 & .901 & [-0.19, 0.22] \\
& full & 910 & 649 & .23 & n/a & n/a & n/a & n/a \\
\rowcolor{lightgray} 
M5 & base & 1,056 & 1,064 & .00 & 0.09 & 1.13 & .257 & [-0.06, 0.24] \\
\rowcolor{lightgray} 
& full & 1,178 & 1,014 & .38 & n/a & n/a & n/a & n/a \\
M6 & \textbf{base*} & 989 & 1,024 & .00 & 0.17 & 2.22 & \textbf{.026} & [0.02, 0.31] \\
& full & 1,217 & 1,089 & .41 & 0.07 & 1.47 & .142 & [-0.03, 0.17] \\
\bottomrule
\vspace{3pt}
\end{tabular}
\begin{tabular}{lcccccccccccccccc}
\toprule
\multicolumn{17}{c}{\textbf{Selected confounders in the full model (notification type P4)}} \\ 
Metric & P1 & P2 & P3 & P5 & P6 & WD & O3 & PM & CO & SO2 & XH & YH & XM & YM & E & M\\ 
\midrule
\rowcolor{lightgray} 
M1 &  &  & \checkmark & ... & ... &  & \checkmark & ... & ... & ... & \checkmark & \checkmark &  & \checkmark & \checkmark & \checkmark \\
M2 &  &  & \checkmark &  & \checkmark & ... & \checkmark & ... & ... & ... & \checkmark & \checkmark &  & \checkmark & \checkmark & \checkmark \\
\rowcolor{lightgray} 
M3 & \checkmark &  & \checkmark &  & \checkmark &  & ... & ... & ... & ... & \checkmark & \checkmark & ... & \checkmark &  & \checkmark \\
M4 &  &  &  & ... &  &  & \checkmark & ... &  & ... & \checkmark & \checkmark &  &  & \checkmark & \checkmark \\
\rowcolor{lightgray} 
M5 &  &  &  &  &  &  & \checkmark &  &  &  & \checkmark & \checkmark &  &  & \checkmark & \checkmark \\
M6 &  &  & \checkmark & \checkmark & \checkmark &  & ... & ... & ... & ... & \checkmark & \checkmark & ... & \checkmark &  & \checkmark \\
\bottomrule
\end{tabular}
\end{table}

\begin{table}[p]
	\caption{Analysis of the metrics (Table~\ref{tab:pn_type}) between two groups: within two hours \textbf{before and after} sending \textbf{type P5 notifications} (\textit{PGH Air Quality Notification: PGH Air Quality Notification AQI has been over 50 for last 2 hrs.}). The \quo{Before} and \quo{After} columns show \quo{median $\pm$ semi-interquartile range}. Symbol $\rho$ is the Spearman's correlation coefficient (effect size). Other notations and abbreviations in this table are explained in Table~\ref{tab:pn_analysis_tp_fn}.}
	\label{tab:pn_analysis_pn5}
	\begin{tabular}{lccccc}
\toprule
\multicolumn{6}{c}{\textbf{Wilcoxon signed-rank test (notification type P5), N=1978}} \\ 
Metric & Before & After & T statistic & p-value & $\rho$ \\ 
\midrule
\rowcolor{lightgray} 
M1 & 1$\pm$1 & 1$\pm$2 & 410,720 & .445 & .46 \\
M2 & 1$\pm$1 & 1$\pm$1 & 401,178 & .415 & .47 \\
\rowcolor{lightgray} 
M3 & 1$\pm$1 & 1$\pm$1 & 382,528 & .162 & .46 \\
M4 & 7$\pm$12 & 8$\pm$14 & 674,426 & \textbf{<.001} & .43 \\
\rowcolor{lightgray} 
M5 & 1$\pm$2 & 2$\pm$2 & 506,584 & \textbf{<.001} & .45 \\
M6 & 4$\pm$4 & 4$\pm$4 & 637,138 & \textbf{<.001} & .53 \\
\bottomrule
\vspace{3pt}
\end{tabular}
\begin{tabular}{llccccccc}
\toprule
\multicolumn{9}{c}{\textbf{Regression analysis (notification type P5)}} \\ 
Metric & Model & $D$ & $P$ & $R^2$ & Coef & Z statistic & p-value & CI \\ 
\midrule
\rowcolor{lightgray} 
M1 & base & 1,585 & 2,246 & .00 & -0.03 & -0.31 & .754 & [-0.21, 0.15] \\
\rowcolor{lightgray} 
& full & 2,730 & 3,405 & .34 & n/a & n/a & n/a & n/a \\
M2 & base & 1,568 & 2,240 & .00 & -0.03 & -0.27 & .784 & [-0.21, 0.16] \\
& full & 2,672 & 3,226 & .35 & n/a & n/a & n/a & n/a \\
\rowcolor{lightgray} 
M3 & base & 3,290 & 3,366 & .00 & 0.04 & 0.79 & .430 & [-0.06, 0.14] \\
\rowcolor{lightgray} 
& full & 3,296 & 3,205 & .37 & 0.06 & 1.39 & .165 & [-0.02, 0.14] \\
M4 & base & 3,559 & 3,884 & .00 & 0.09 & 1.41 & .157 & [-0.04, 0.22] \\
& full & 2,725 & 1,882 & .23 & 0.08 & 1.25 & .210 & [-0.04, 0.21] \\
\rowcolor{lightgray} 
M5 & \textbf{base***} & 3,088 & 3,473 & .00 & 0.16 & 3.38 & \textbf{<.001} & [0.07, 0.26] \\
\rowcolor{lightgray} 
& \textbf{full***} & 3,376 & 3,186 & .38 & 0.13 & 3.56 & \textbf{<.001} & [0.06, 0.20] \\
M6 & \textbf{base***} & 3,137 & 3,788 & .01 & 0.20 & 4.20 & \textbf{<.001} & [0.10, 0.29] \\
& \textbf{full***} & 3,716 & 3,975 & .42 & 0.12 & 3.87 & \textbf{<.001} & [0.06, 0.19] \\
\bottomrule
\vspace{3pt}
\end{tabular}
\begin{tabular}{lcccccccccccccccc}
\toprule
\multicolumn{17}{c}{\textbf{Selected confounders in the full model (notification type P5)}} \\ 
Metric & P1 & P2 & P3 & P4 & P6 & WD & O3 & PM & CO & SO2 & XH & YH & XM & YM & E & M\\ 
\midrule
\rowcolor{lightgray} 
M1 & \checkmark &  &  &  & ... &  & \checkmark & ... &  & \checkmark & \checkmark & \checkmark & ... & \checkmark & ... & \checkmark \\
M2 & \checkmark &  &  &  & ... &  & \checkmark & ... &  & \checkmark & \checkmark & \checkmark & ... & \checkmark & ... & \checkmark \\
\rowcolor{lightgray} 
M3 & \checkmark &  & \checkmark &  & \checkmark &  & ... &  & ... & \checkmark & \checkmark & \checkmark &  & \checkmark &  & \checkmark \\
M4 &  &  &  & ... &  &  &  & ... & ... & \checkmark & \checkmark & \checkmark &  & \checkmark & ... & \checkmark \\
\rowcolor{lightgray} 
M5 & \checkmark &  & \checkmark & ... & \checkmark &  &  & ... & ... & \checkmark & \checkmark & \checkmark &  & \checkmark & \checkmark & \checkmark \\
M6 & \checkmark &  & \checkmark & \checkmark & \checkmark & ... & ... & ... & ... & \checkmark & \checkmark & \checkmark &  & \checkmark & \checkmark & \checkmark \\
\bottomrule
\end{tabular}
\end{table}

\begin{table}[p]
	\caption{Analysis of the metrics (Table~\ref{tab:pn_type}) between two groups: within two hours \textbf{before and after} sending \textbf{type P6 notifications} (\textit{Smell Report Summary: 53 smell reports were submitted today.}). The \quo{Before} and \quo{After} columns show \quo{median $\pm$ semi-interquartile range}. Symbol $\rho$ is the Spearman's correlation coefficient (effect size). Other notations and abbreviations in this table are explained in Table~\ref{tab:pn_analysis_tp_fn}.}
	\label{tab:pn_analysis_pn6}
	\begin{tabular}{lccccc}
\toprule
\multicolumn{6}{c}{\textbf{Wilcoxon signed-rank test (notification type P6), N=961}} \\ 
Metric & Before & After & T statistic & p-value & $\rho$ \\ 
\midrule
\rowcolor{lightgray} 
M1 & 1$\pm$2 & 2$\pm$1 & 102,821 & \textbf{<.001} & .33 \\
M2 & 1$\pm$1 & 2$\pm$1 & 99,646 & \textbf{<.001} & .32 \\
\rowcolor{lightgray} 
M3 & 1$\pm$1 & 2$\pm$1 & 96,678 & \textbf{<.001} & .33 \\
M4 & 13$\pm$14 & 30$\pm$24 & 80,646 & \textbf{<.001} & .48 \\
\rowcolor{lightgray} 
M5 & 2$\pm$2 & 5$\pm$2 & 36,255 & \textbf{<.001} & .52 \\
M6 & 6$\pm$4 & 12$\pm$6 & 27,412 & \textbf{<.001} & .57 \\
\bottomrule
\vspace{3pt}
\end{tabular}
\begin{tabular}{llccccccc}
\toprule
\multicolumn{9}{c}{\textbf{Regression analysis (notification type P6)}} \\ 
Metric & Model & $D$ & $P$ & $R^2$ & Coef & Z statistic & p-value & CI \\ 
\midrule
\rowcolor{lightgray} 
M1 & \textbf{base**} & 1,678 & 1,988 & .01 & 0.18 & 3.23 & \textbf{.001} & [0.07, 0.29] \\
\rowcolor{lightgray} 
& \textbf{full***} & 1,946 & 1,928 & .25 & 0.18 & 4.12 & \textbf{<.001} & [0.10, 0.27] \\
M2 & \textbf{base***} & 1,630 & 1,905 & .01 & 0.19 & 3.36 & \textbf{<.001} & [0.08, 0.30] \\
& \textbf{full***} & 1,899 & 1,881 & .24 & 0.19 & 4.26 & \textbf{<.001} & [0.10, 0.28] \\
\rowcolor{lightgray} 
M3 & \textbf{base***} & 1,565 & 1,403 & .01 & 0.20 & 3.77 & \textbf{<.001} & [0.10, 0.30] \\
\rowcolor{lightgray} 
& \textbf{full***} & 1,792 & 1,540 & .29 & 0.21 & 5.29 & \textbf{<.001} & [0.14, 0.29] \\
M4 & \textbf{base***} & 1,751 & 1,753 & .06 & 0.70 & 10.79 & \textbf{<.001} & [0.57, 0.83] \\
& \textbf{full***} & 1,263 & 896 & .32 & 0.68 & 10.49 & \textbf{<.001} & [0.56, 0.81] \\
\rowcolor{lightgray} 
M5 & \textbf{base***} & 1,559 & 1,797 & .12 & 0.67 & 14.35 & \textbf{<.001} & [0.58, 0.76] \\
\rowcolor{lightgray} 
& \textbf{full***} & 1,809 & 1,581 & .52 & 0.68 & 21.55 & \textbf{<.001} & [0.62, 0.75] \\
M6 & \textbf{base***} & 1,091 & 1,668 & .15 & 0.67 & 14.09 & \textbf{<.001} & [0.58, 0.77] \\
& \textbf{full***} & 1,862 & 1,920 & .58 & 0.70 & 27.27 & \textbf{<.001} & [0.65, 0.75] \\
\bottomrule
\vspace{3pt}
\end{tabular}
\begin{tabular}{lccccccccccccc}
\toprule
\multicolumn{14}{c}{\textbf{Selected confounders in the full model (notification type P6)}} \\ 
Metric & P4 & P5 & WD & O3 & PM & CO & SO2 & XH & YH & XM & YM & E & M\\ 
\midrule
\rowcolor{lightgray} 
M1 & \checkmark &  &  & \checkmark & ... & ... & ... &  &  & \checkmark & \checkmark & \checkmark & \checkmark \\
M2 & \checkmark &  &  & \checkmark & ... & ... & ... &  &  & ... & \checkmark & \checkmark & \checkmark \\
\rowcolor{lightgray} 
M3 & \checkmark &  & ... & ... & \checkmark &  &  & \checkmark &  & ... & ... & \checkmark & \checkmark \\
M4 &  &  &  &  & ... &  &  &  &  &  &  & \checkmark & \checkmark \\
\rowcolor{lightgray} 
M5 &  &  & \checkmark & ... &  &  & ... & \checkmark &  & \checkmark & ... & \checkmark & \checkmark \\
M6 &  &  & \checkmark & ... & ... & ... & ... & \checkmark &  & \checkmark & \checkmark & \checkmark & \checkmark \\
\bottomrule
\end{tabular}
\end{table}

\newpage

\subsection{Materials for the Survey Study}\label{ap:survey_materials}

The following pages contain materials used for the survey study in section~\ref{sec:survey_study}.

\newpage

\foreach \n in {1,...,9}{
	\begin{figure}
			\centering
			\includegraphics[page=1,width=1\columnwidth]{file/survey-\n.pdf}
	\end{figure}
}
	
\end{document}